\documentclass[twocolumn,superscriptaddress,amsmath,aps,longbibliography,nofootinbib]{revtex4-1}

\usepackage{amsmath,amsfonts}
\usepackage{dcolumn}
\usepackage{bm}
\usepackage{color}
\usepackage{multirow}
\usepackage{scrextend}

\usepackage{graphicx}
\usepackage{hyperref}
\usepackage{float} 
\usepackage[utf8x]{inputenc}

\usepackage{natbib}
\begin{document}


\title{Strong-coupling superconductivity of SrIr$_2$ and SrRh$_2$:  Phonon engineering of metallic Ir and Rh}



\author{Sylwia Gutowska}
\email{gutowska@agh.edu.pl}
\affiliation{Faculty of Physics and Applied Computer Science,
AGH University of Science and Technology, Aleja Mickiewicza 30, 30-059 Krakow, Poland}
\author{Karolina Górnicka}
\affiliation{Faculty of Applied Physics and Mathematics and Advanced Materials Centre, Gdansk University of Technology,
ul. Narutowicza 11/12, 80-233 Gdańsk, Poland}
\author{Pawe\l{} W\'ojcik}
\affiliation{Faculty of Physics and Applied Computer Science,
AGH University of Science and Technology, Aleja Mickiewicza 30, 30-059 Krakow, Poland}
\author{Tomasz Klimczuk}
\affiliation{Faculty of Applied Physics and Mathematics and Advanced Materials Centre, Gdansk University of Technology,
ul. Narutowicza 11/12, 80-233 Gdańsk, Poland}
\author{Bartlomiej Wiendlocha}
\email{wiendlocha@agh.edu.pl}
\affiliation{Faculty of Physics and Applied Computer Science,
AGH University of Science and Technology, Aleja Mickiewicza 30, 30-059 Krakow, Poland}


\date{\today}

\begin{abstract}
Experimental and theoretical studies on superconductivity in SrIr$_2$ and SrRh$_2$ Laves phases are presented. The measured resistivity, heat capacity, and magnetic susceptibility confirm the superconductivity of these compounds with $T_c$ = 6.07\,K and 5.41\,K, respectively. 
Electronic structure calculations show that the Fermi surface is mostly contributed by 
5$d$ (4$d$) electrons of Ir (Rh), with Sr atoms playing the role of electron donors. 
The effect of the spin-orbit coupling is analyzed and found to be important in both materials.
Lattice dynamics and electron-phonon coupling (EPC) are studied and the strong electron-phonon interaction is found, contributed mostly by the low-frequency Ir and Rh vibrations.
The enhancement of EPC when compared to weakly-coupled metallic Ir and Rh is explained by the strong modifications in the propagation of phonons in the network of Ir (Rh) tetrahedrons, which are the building blocks of the Laves phase, and originate from the metallic fcc structures of elemental iridium and rhodium.
\end{abstract}


\maketitle

\section{Introduction}

The widespread family of AB$_2$ compounds, forming the so-called Laves phases \cite{laves1,laves2,laves3}
is characterized by the C15-type (cubic {fcc} MgCu$_2$-type) or C14-type (hexagonal MgZn$_2$- or MgNi$_2$- type) crystal structures. Despite the similarities in their structures, Laves phases exhibit a multitude of physical properties including magnetoelastic phase transitions,  superconductivity, high hardness, corrosion resistance, ability to storage hydrogen, and on this basis multiple applications in industry have been proposed~\cite{laves-review,laves-review2}. 
Moreover, the question why this specific type of crystal structure is so often found among intermetallic compounds \cite{laves-compounds,laves-compounds2} and alloys \cite{laves-alloys}, including high entropy alloys \cite{laves-hea,laves-hea2}, is frequently raised.

Superconductivity in C15-type Laves phases of SrIr$_2$ and SrRh$_2$ was first reported by Matthias and Corenzwit~\cite{matthias} with $T_c = 5.7$ K and 6.2 K, respectively. Very recently, Horie {\it et al.} \cite{srir2-horie} reported the superconducting state parameters for SrIr$_2$ with $T_c = 5.90$\,K, whereas Gong {\it et al.} \cite{gong-srrh2-barh2}
performed measurements for SrRh$_2$ indicating $T_c = 5.40$\,K.  
Those results show that both compounds are moderately or strongly coupled type-II BCS superconductors, and the order of $T_c$'s is opposite to that reported by Matthias and Corenzwit~\cite{matthias}, with larger $T_c$ for the Ir-based compound.  Further on, Yang {\it et al.}~\cite{srir2-yang} reported even higher $T_c = 6.6$\, K in SrIr$_2$ and suggested a non-BCS like behavior of superconductivity under pressure.

In this work, we extend the analysis of superconductivity and its relation to the crystal structure in the two aforementioned Laves phases, SrIr$_2$ and SrRh$_2$, by means of experimental and theoretical studies.
Resistivity, heat capacity, and magnetic susceptibility are  measured down to 1.7 K and confirm bulk superconductivity with $T_c = 6.07$~K in SrIr$_2$ and $T_c = 5.41$~K in SrRh$_2$.
Electronic structure, lattice dynamics, and the electron-phonon interaction are studied using the density functional theory (DFT), and the spin-orbit coupling (SOC) effects are discussed.
Theoretical results show that both compounds are strongly coupled superconductors, with the electron-phonon coupling (EPC) parameter $\lambda \sim 1$.

As listed above, 
experimental studies of superconductivity in SrIr$_2$ and SrRh$_2$ indicate that their critical temperature may vary depending on sample preparation details, and are in the range 5.7--6.6\,K (SrIr$_2$) \cite{srir2-horie,srir2-yang,matthias} and 5.4--6.2\,K (SrRh$_2$) \cite{gong-srrh2-barh2,matthias}. 
Such differences are not unusual, however, importantly for the understanding of superconductivity in these materials, the critical temperatures measured in this work (6.07\,K and 5.41\,K respectively), together with the presented theoretical analysis show that SrRh$_2$ is a superconductor with a lower critical temperature than SrIr$_2$. 

\begin{figure*}[t]
\includegraphics[width=0.95\textwidth]{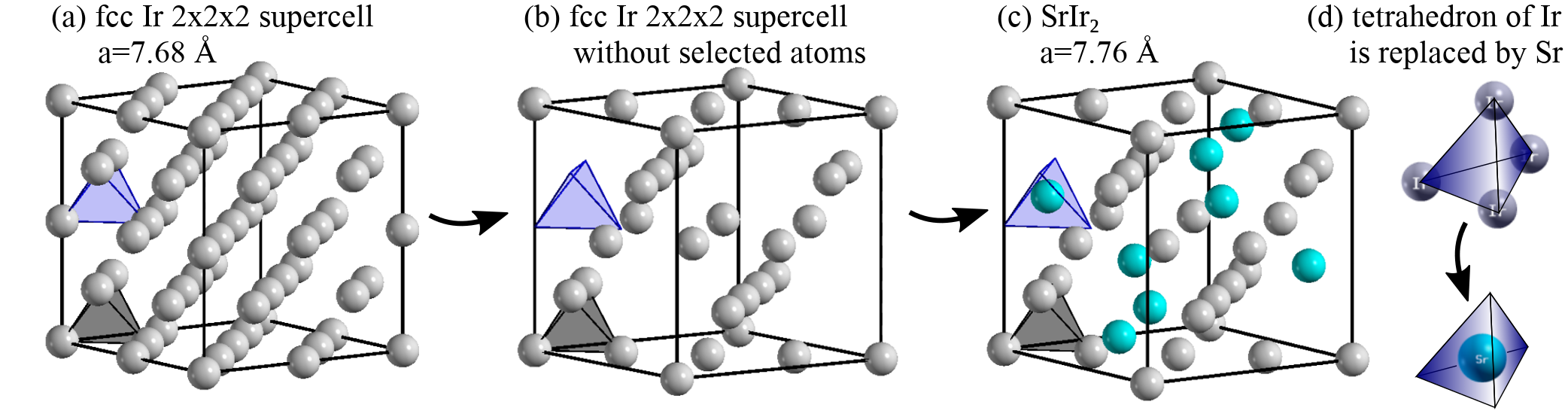}
\caption{The crystal structure of SrIr$_2$ shown as a modified structure of elemental metallic Ir: (a) 2x2x2 supercell of {fcc} Ir with two tetrahedrons built of Ir, marked with blue and black color. The whole structure consist of two sublattices of blue and black tetrahedrons; (b) the atoms associated with blue tetrahedrons are removed from the cell; (c) 
the center of the removed tetrahedrons is filled with Sr atom, while the black tetrahedron sublattice remains unchanged; (d) shows schematic replacement of Ir$_4$ tetrahedron by one Sr atom. In panel (a) only one of total eight Ir$_4$ tetrahedrons, which are replaced by Sr, is marked, for more details and a relation between the primitive cells see Fig. S2 in Ref. \cite{supplemental}.}\label{fig:structure}
\end{figure*}

Structurally, the Laves phases of Sr$M_2$ ($M = $ Ir and Rh) may be viewed as the modified metallic { fcc} structures of iridium and rhodium, whose face-centered cubic elemental cells are formed from the tetrahedrons of $M$-atoms, stacked in a dense-packed structure. 
To form the Laves phase, a large Sr atom is inserted to the structure where it substitutes every second tetrahedron of $M$ atoms (see, Fig.~\ref{fig:structure}), leaving the remaining unit cell geometry unchanged, only slightly  (about 1\%) expanding the lattice.
Thus, what mainly caught our attention was how this $M_4 \rightarrow$ Sr substitution changes the very-low-$T_c$ superconductors ($T_c$ of 0.14~K in Ir \cite{iridium-superconductivity} and 0.3~mK in Rh  \cite{rhodium-superconductivity}) 
into the strongly-coupled superconductors with a decent $T_c$, especially that in both the electronic and phonon structures of Sr$M_2$ we find similarities to the elemental { fcc} crystals of $M$.
What we found may be called the "phonon engineering" effect, as the strong enhancement of both the electron-phonon interaction and $T_c$ of Sr$M_2$, in comparison to Ir and Rh, is caused
by the substantial  lowering of the frequency of several phonon modes, propagating in the network of $M$ tetrahedrons spaced by Sr atoms. These changes are directly related to the replacing of half of dense-packed  tetrahedrons by Sr.

\section{Materials and Methods}
Polycrystalline samples of SrIr$_2$ and SrRh$_2$ were prepared by a two-step solid state reaction method from the required high-purity elements i.e., Sr-pieces (4N, Onyxmet), Ir-powder (4N, Mennica-Metale, Poland) and Rh-powder (3N8, Mennica-Metale, Poland). An excess of strontium (25\%) was added in order to compensate for its loss during the synthesis. The following manipulations were performed in a protective Ar-atmosphere glove box system (p(O$_2$) $<$ 0.5 ppm). The starting materials were mixed, put into the alumina crucible, and then sealed inside an Ar-filled quartz tube. The ampule was heated to 700$^{\circ}$C, then slowly heated to 790$^{\circ}$C at a rate of 10$^{\circ}$C/h, held at that temperature for 6h and then slowly cooled (20$^{\circ}$C/h) and kept at 700$^{\circ}$C for 2\,h. The as-prepared material was reground well and pressed into a pellet. The samples were then enclosed inside a quartz tube and annealed at 800$^{\circ}$C for 12 hours. Products were dense and black in color. To avoid possible decomposition, the samples were kept inside the glove box until characterization.
The phase purity and structure of the samples were checked by room-temperature powder x-ray diffraction (pXRD) characterization carried out on a Bruker D2 PHASER diffractometer with Cu K$_\alpha$ radiation ($\lambda$ = 1.5406\,\AA). The data were analyzed by the LeBail method implemented in the {\sc diffrac.suite.topas}. Magnetization measurements were performed using a  Quantum Design EverCool II Physical Property Measurement System (PPMS) with a vibrating sample magnetometer (VSM) function. Both zero-field-cooled (ZFC) and field-cooled (FC) data were collected from 1.7 to 6.5\,K under an applied field of 10\,Oe. The magnetization was also measured at various temperatures in the superconducting state as a function of the applied field. 
Resistivity and heat capacity measurements were performed on a PPMS Evercool II. Specific heat was measured in zero field and field up to 6\,T, using the two-$\tau$ time-relaxation method. The resistivity was determined using a standard four-probe technique, with four 57-$\mu$m-diameter platinum wire leads attached to the flat polished sample surface using conductive silver epoxy (Epotek H20E).

\begin{figure*}[t]
\includegraphics[width=0.85\textwidth]{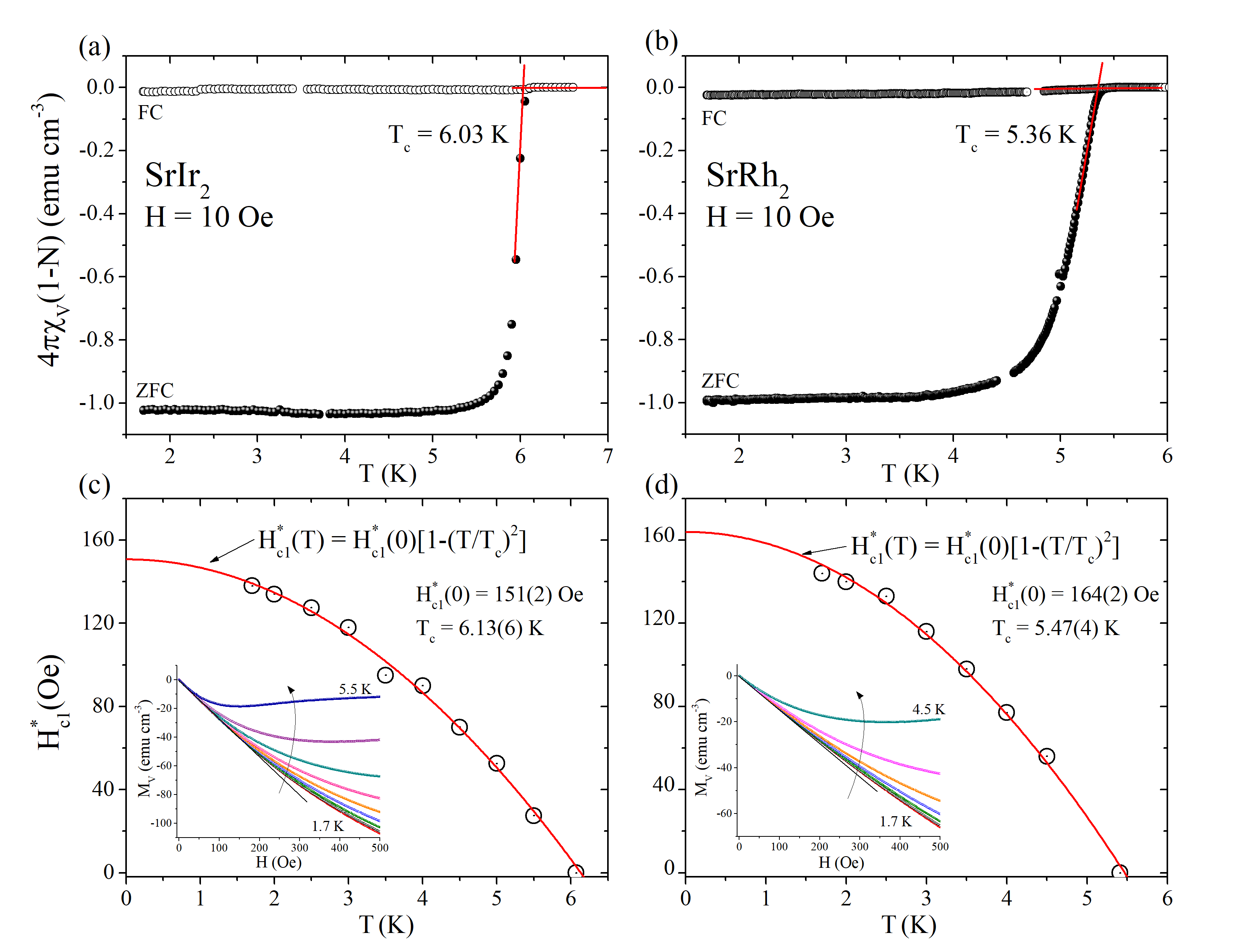}
\caption{Zero-field-cooled (open circles) and field-cooled (full circles) temperature-dependent magnetic susceptibility data in $H = 10$\,Oe for SrIr$_2$ (a) and SrRh$_2$ (b).
The temperature dependence of the lower critical fields for (c) SrIr$_2$; and (d) SrRh$_2$. Insets in (c-d) show the field-dependent magnetization curves $M_V(H)$ taken at different temperatures.}\label{fig:exp:magnetic-properties}
\end{figure*}

Theoretical calculations of the electronic structure, phonons, and the electron-phonon interaction function were done using the plane-wave pseudopotential method, implemented in the {\sc Quantum espresso} (QE)~\cite{qe,qe2} package. Ultrasoft pseudopotentials were used for all atoms with the exchange-correlation effects included within the Perdew-Burke-Ernzerhof generalized gradient approximation (GGA) scheme~\cite{pbe}. The pseudopotentials have been generated with the help of QE, using input files \footnote{The input files Rh-(rel-)pbe-rkkjus.in, Ir-(rel-)pbe-rkkjus.in and Sr-(rel-)pbe-rkkjus.in have been used} from {\sc PSlibrary} package \cite{pps}. 
To investigate the effect of the spin-orbit coupling, both the scalar-relativistic and fully-relativistic calculations were performed. The electronic structure was calculated on a grid of $12^{3}$ k-points for the self-consistent cycle and $24^{3}$  and $48^{3}$ for density of states (DOS) and Fermi surface (FS) calculations. Next, the phonons were calculated with the help of density functional perturbation theory (DFPT) \cite{dfpt} on the grid of 6$^{3}$ q-points, which results in the number of 16 independent dynamical matrices to be calculated. Electron-phonon interaction function was calculated on the basis of the obtained phonon structure with the integrals over Fermi surface calculated with double-delta smearing technique with the smearing parameter $\sigma=0.02$ Ry, and on the basis of the electronic structure calculated on the mesh of $24^3$ k-points.

\section{Experimental studies}

\begin{figure*}[t]
\includegraphics[width=0.85\textwidth]{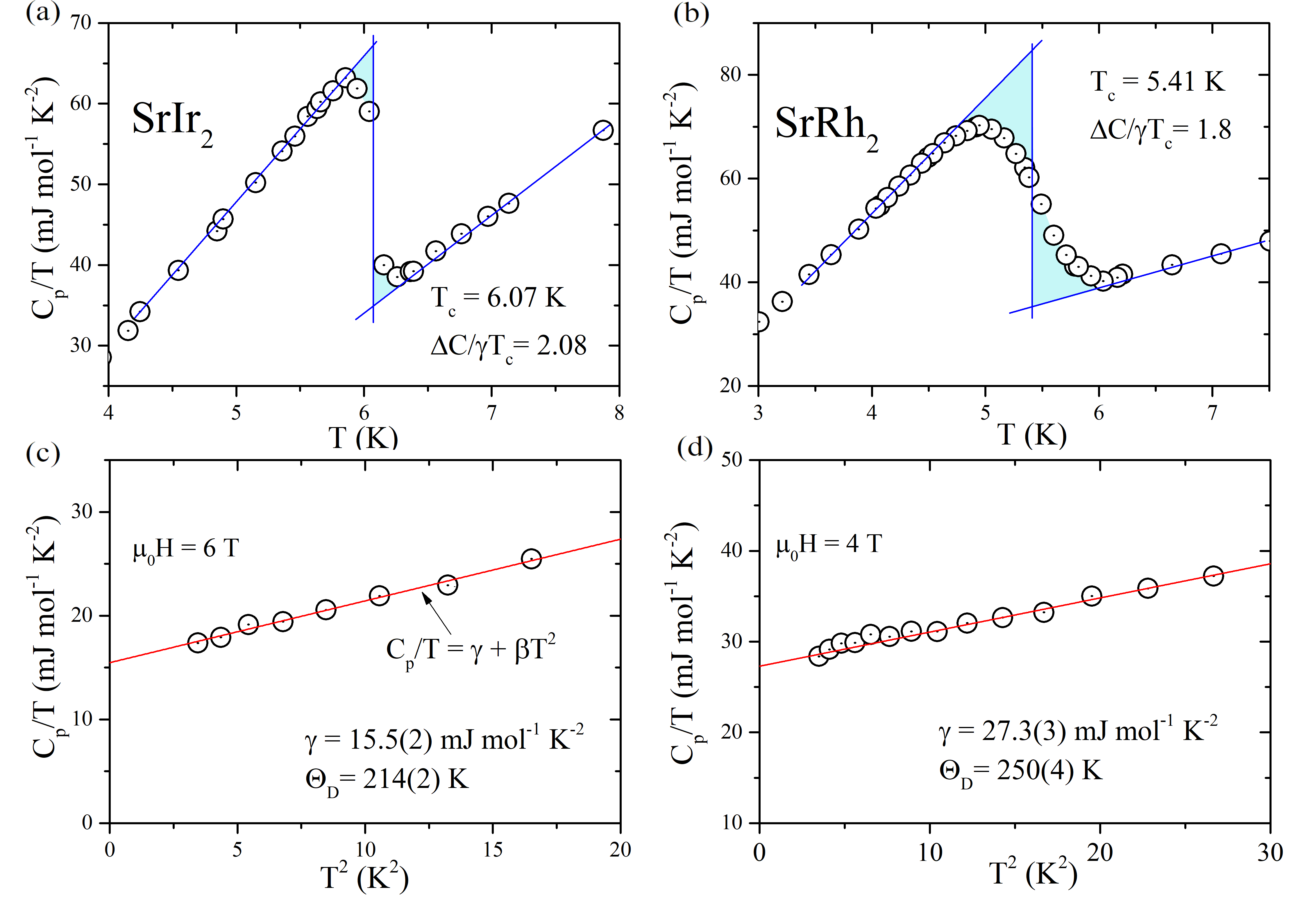}
\caption{Zero-field specific heat divided by temperature ($C_p/T$ ) versus temperature for (a) SrIr$_2$; and (b) SrRh$_2$.
$C_p/T$ versus $T^2$ measured in 6~T and 4~T field (in the normal state) for (c) SrIr$_2$; and (d) SrRh$_2$, fitted to $C_p/T = \gamma + \beta T^2$.}\label{fig:exp:heat-capacity}
\end{figure*}

The room temperature powder XRD patterns (Fig. S1 in the Supplemental Material~\cite{supplemental}) for Sr$M_2$ ($M$ = Ir or Rh) confirm that both compounds crystallize in a cubic Laves-phase type structure (space group \makebox{$Fd$-$3m$,} no. 227), shown in Fig.~\ref{fig:structure}(c). The LeBail refinement yielded the lattice parameters $a=7.7932(1)$\,\AA~(SrIr$_2$) and $a= 7.7069(4)$\,\AA~(SrRh$_2$). These values are in good agreement with those reported in the literature \cite{srir2-srrh2-lattice-constant}. The Sr and Ir atoms occupy $8b$ and $16c$ Wyckhoff positions.
Superconductivity of both compounds was confirmed by magnetic measurements as shown in Fig.~\ref{fig:exp:magnetic-properties}. The zero-field-cooled (ZFC) and field-cooled (FC) volume magnetic susceptibility (defined as $\chi_V=\frac{M_V}{H}$ where $M_V$ is the volume magnetization and $H$ is the applied magnetic field) versus temperature, measured under 10\,Oe magnetic field, is shown in Fig.~\ref{fig:exp:magnetic-properties}(a) for SrIr$_2$ and Fig.~\ref{fig:exp:magnetic-properties}(b) for SrRh$_2$.
The observed strong diamagnetic signal below the transition temperature $T_c$ = 6.03\,K for SrIr$_2$ and $T_c$ = 5.36\,K for SrRh$_2$ confirms the appearance of superconductivity in both compounds. The critical superconducting temperature was estimated as the intersection between two lines marked in red [see Fig.~\ref{fig:exp:magnetic-properties}(a) and (b)]. 
The obtained $T_c$ for SrIr$_2$ is slightly higher than the value reported previously by Horie {\it et al.} (5.9\,K) \cite{srir2-horie},  and lower than 6.6\,K reported by Yang  {\it et al.}, while for SrRh$_2$ $T_c$ is almost identical with that reported by Gong {\it et al.} \cite{gong-srrh2-barh2}. The experimental data were corrected for a demagnetization factor $N$ estimated from an isothermal $M_V(H)$ curves, as explained below. At 1.7\,K the $4\pi\chi_V V(1-N)$ (ZFC) value approaches \makebox{-1} indicating that the Meissner volume fraction is 100$\%$. The FC signal is much weaker compared to the ZFC signal, usually resulting from strong flux trapping, as is typically observed in polycrystalline samples.

The magnetic isotherms measured over a range of temperatures below $T_c$ are presented in the inset of Fig.~\ref{fig:exp:magnetic-properties}(c) for SrIr$_2$ and Fig.~\ref{fig:exp:magnetic-properties}(d) for SrRh$_2$. Assuming that the initial linear response to the field is perfectly diamagnetic, we obtain a demagnetization factor $N$, which is 0.71 for the Ir variant and 0.45 for the Rh variant. For an analysis of the lower critical field ($H_{c_1}^*$) the point of deviation from the full Meissner state was estimated for each temperature. To precisely calculate this point, we follow the methodology described elsewhere \cite{critical-h,critical-h2}. The temperature dependence of $H_{c_1}^*$ is plotted in the main panel of Fig.~\ref{fig:exp:magnetic-properties}(c) and Fig.~\ref{fig:exp:magnetic-properties}(d) for SrIr$_2$ and SrRh$_2$, respectively. An additional point for $H=0$ is a zero field transition temperature taken from the heat capacity measurement. The values determined for both samples are fitted using the Ginzburg-Landau equation \cite{SC_book2}

\begin{equation}
H^*_{c_1}(T)=H^*_{c_1}(0)\Big[1-\Big(\frac{T}{T_c}\Big)^2\Big]
\end{equation}
where $H^*_{c_1}(0)$ is the lower critical field at 0\,K. The quadratic expression fits the data very well and yields $H^*_{c_1}(0)=151(2)$\,Oe and $T_c=6.13(6)$\,K for the Ir variant, and $H^*_{c_1}(0)=164(2)$\,Oe and $T_c=5.47(4)$\,K for the Rh variant. Correcting for the demagnetization factor, the lower critical field at 0\,K is calculated (by formula $H_{c_1}=\frac{H_{c_1}^*(0)}{1-N}$) to be $H_{c_1}(0) = 522$\,Oe ($52.2$\,mT) for SrIr$_2$ and $H_{c_1}(0)= 298$\,Oe ($29.8$\,mT) for SrRh$_2$.  Estimated values are larger than the 
previously reported by Horie {\it et al.} \cite{srir2-horie} for SrIr$_2$ (10.1\,mT) and Gong {\it et al.} \cite{gong-srrh2-barh2} for SrRh$_2$ (10.1(3)\,mT). The discrepancy may be due to the demagnetization correction that was applied in this work.

\begin{figure*}[ht]
\includegraphics[width=1.00\textwidth]{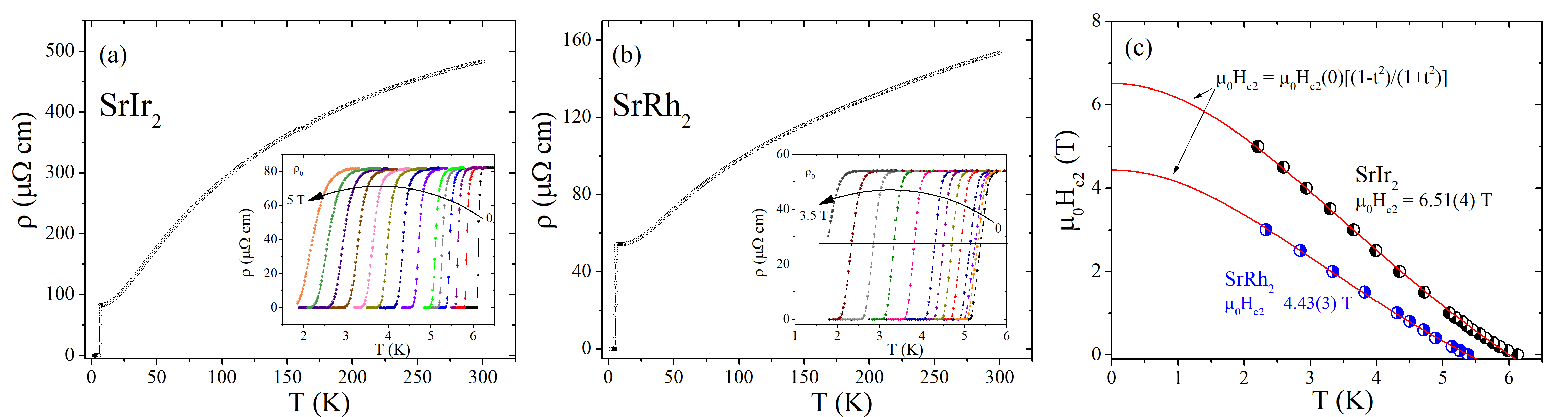}
\caption{The electrical resistivity of (a) SrIr$_2$; and (b) SrRh$_2$ versus temperature, measured in zero applied magnetic field. The insets show the expanded
plot of $\rho(T)$ in the vicinity of the superconducting transition
for different values of $H$;
(c) Temperature dependence of the upper critical field, determined from the electrical resistivity data.}\label{fig:exp:resistivity}
\end{figure*}

The characterization of the superconducting transition by specific heat measurements is summarized in Fig.~\ref{fig:exp:heat-capacity}(a) and \ref{fig:exp:heat-capacity}(c) for SrIr$_2$, and Fig.~\ref{fig:exp:heat-capacity}(b) and \ref{fig:exp:heat-capacity}(d) for SrRh$_2$. 
Panels (a) and (b) show the zero-field data plotted as $C_p/T$ versus $T$. The bulk superconductivity in both compounds is manifested by a  large anomaly at $6.07$\,K for SrIr$_2$ and $5.41$\,K for SrRh$_2$, in agreement with the magnetization data presented above. 
In contrary to a sharp heat capacity jump for SrIr$_2$, a rather broad transition for SrRh$_2$ is observed. This may be caused by the sample inhomogeneity, supported by a lower residual resistivity ratio (RRR) value, discussed below.
Experimental data measured under the applied magnetic field [panels (c) and (d) in Fig.~\ref{fig:exp:heat-capacity}], represents the normal state specific heat of both compounds. 
The data plotted as $C_p/T$ versus $T^2$ can be fitted using the formula $\frac{C_p}{T}=\gamma + \beta T^2$, where the first and the second terms are the electronic specific heat coefficient and phonon contribution, respectively. The fit yields $\gamma= 15.5(2)$\,mJ$\cdot$mol$^{-1}$K$^{-2}$ and $\beta= 0.595(1)$\,mJ$\cdot$mol$^{-1}$K$^{-4}$ for SrIr$_2$ and $\gamma= 27.3(3)$\,mJ$\cdot$mol$^{-1}$K$^{-2}$ and $\beta=0.376(4)$\,mJ$\cdot$mol$^{-1}$K$^{-4}$ for SrRh$_2$. The Debye temperature ($\Theta_D$) can be then calculated through the  relationship $\Theta_D=\Big(\frac{12\pi^4}{5\beta}nR\Big)^{\frac{1}{3}}$, where $R=8.31$\,J$\cdot$mol$^{-1}$K$^{-1}$ is the ideal gas constant,  and $n = 3$ is the number of atoms per formula unit. The resulting values of $\Theta_D$ are $214(2)$\,K for SrIr$_2$ and $250(4)$\,K for SrRh$_2$, however, one has to remember that for such a complex crystal structure, with six atoms in the primitive cell, the phonon spectrum will be far from a simple Debye model, as we will see below.
The calculated value of Debye temperature for SrIr$_2$ is slightly higher than reported previously (180\,K \cite{srir2-horie}) however, it should be noted that the value of $\beta$, and hence $\Theta_D$, is strongly affected by the temperature range employed for the fit.

As commonly practised with these parameters, the EPC constant $\lambda$ can be estimated from the McMillan’s equation \cite{mcmillan}
\begin{equation}\label{eq:mcmillan}
\lambda = \frac{1.04+\mu^*\ln\Big(\frac{\Theta_D}{1.45T_c}\Big)}
{(1-0.62\mu^*)\ln\Big(\frac{\Theta_D}{1.45T_c}\Big)-1.04}
\end{equation}
where $\mu^*$ is the repulsive screened Coulomb pseudopotential parameter, usually taken as $\mu^* = 0.13$ \cite{mcmillan}. In that case, one obtains  $\lambda= 0.77$ for SrIr$_2$ and  $\lambda= 0.70$ for SrRh$_2$, suggesting that both compounds are moderately coupled superconductors. Furthermore, for both compounds, the normalized specific heat jump ($\Delta C/\gamma T_c$), equal to $2.08$ for SrIr$_2$ and $1.8$ for SrRh$_2$  exceeds the weak-coupling BCS value of $1.43$, similarly to what was reported in Refs.~\cite{srir2-horie,gong-srrh2-barh2}.
For such a case, we may apply the approximate formula \cite{marsiglio-carbotte} for the specific heat jump in strongly-coupled superconductors: 
\begin{equation}\label{eq:marsiglio}
\frac{\Delta C_p}{\gamma T_c}=1.43\left[1+53\left(\frac{T_c}{\langle\omega_{\rm log}^{\alpha^2F}\rangle}\right)^2\ln\left(\frac{\langle\omega_{\rm log}^{\alpha^2F}\rangle}{3T_c}\right)\right]
\end{equation}
which allows to estimate the logarithmic average phonon frequency $\langle\omega_{\rm log}^{\alpha^2F}\rangle$ weighted by the electron-phonon interaction (see Eq.~\ref{eq:omlog2}). We obtain $\langle\omega_{\rm log}^{\alpha^2F}\rangle = 79$\,K for SrIr$_2$ and  $\langle\omega_{\rm log}^{\alpha^2F}\rangle = 104$\,K for SrRh$_2$. 
Having $\langle\omega_{\rm log}^{\alpha^2F}\rangle$ we may avoid relying on the Debye approximation for the phonon spectrum 
and calculate $\lambda$ using the Allen-Dynes formula for $T_c$ \cite{allen-dynes}: 
\begin{equation}
T_c=\frac{\langle\omega_{\rm log}^{\alpha^2F}\rangle}{1.2}
\exp\left[ 
\frac{-1.04(1+\lambda)}{\lambda-\mu^*(1+0.62\lambda)}\right].
\label{eq:tc}
\end{equation}
Taking $\mu^*=0.13$ the estimated $\lambda$ are $1.17$ for SrIr$_2$ and $0.93$ for SrRh$_2$, and are larger than obtained from the inverted McMillan equation, pointing to the strong coupling regime, in agreement with the enhanced value of $\frac{\Delta C_p}{\gamma T_c}$. These values will be used as the experimental estimates of $\lambda$ in the remaining part of this work.

\begin{table}[b]
\caption{The superconducting properties of SrIr$_2$ and SrRh$_2$ extracted from experimental data. Additionally the critical temperature of SrIr$_2$ equal to $T_c=6.6$\,K was measured in Ref. \cite{srir2-yang}.}
\label{tab:exp}
\begin{center}
\begin{ruledtabular}
\begin{tabular}{l c c c c c c c c}
& 
Unit &
SrIr$_2$ \footnote{\label{thiswork}This work}&
SrIr$_2$ \footnote{\label{horie}\citet*{srir2-horie}}&
SrRh$_2$ \footref{thiswork} &
SrRh$_2$ \footnote{\label{gong}\citet*{gong-srrh2-barh2}} \\
\hline
$T_c$ 			   & K   & 6.07 & 5.90 & 5.41 & 5.40 \\
$\mu_0H_{c_2}(0)$ & T    & 6.51(4)  & 5.90 & 4.43(3) &4.01(5) \\
$\mu_0H_{c_1}(0)$ & mT   & 52.2 & 10.1 & 29.8  & 10.1(3)\\
$\lambda$   & ---  & 1.17 & 0.84 & 0.93 & 0.71 \\
$\xi_{GL}(0)$     & \AA & 71 & 75 & 86 & 91\\
$\lambda_{GL}(0)$ & \AA & 890  & 2370 & 1210 & 2291 \\
$\kappa$          & ---  &12.53  & 31.7  & 14.02 & 23.3(2) \\
$\gamma$          & mJ$\cdot$mol$^{-1}\cdot$K$^{-2}$ & 15.5(2) & 11.9 & 27.3(3) &22.9(3) \\
$\beta$           & mJ$\cdot$mol$^{-1}\cdot$K$^{-4}$ & 0.595(1)& 0.98 & 0.376(4)  & 0.44(3)\\
$\Theta_D$        & K    & 214(2) & 180 & 250(4) & 237(5) \\
RRR               & ---  & 5.9 & ---   & 2.9 & 2.4   \\
$\Delta C_p/(\gamma T_c)$& ---    & 2.08  & 1.71 & 1.80 & 1.78(2) \\
\end{tabular}

\end{ruledtabular}
\end{center}
\end{table}

Figure~\ref{fig:exp:resistivity}(a) and (b) show the temperature-dependent electrical resistivity $\rho(T)$ from $1.8$\,K to $300$\,K for SrIr$_2$ and SrRh$_2$, respectively. For both compounds, the $\rho(T)$ curve in the normal state exhibits typical metallic behavior ($\frac{\text{d}\rho}{\text{d}T} > 0$). The residual resistivity ratio RRR is found to be $\frac{\rho(300)}{\rho(10)} = 5.9$ for the Ir analog and $2.9$ for the Rh analog. These values can be attributed to the polycrystalline nature of our samples. 
At low temperatures,  superconductivity is manifested by a sharp drop in resistivity, down to $\rho = 0$ value. The critical temperature from the resistivity measurements is $T_c = 6.14$\,K for SrIr$_2$ and $T_c = 5.4$\,K for SrRh$_2$, where $T_c$ is defined as $50\%$ decrease of the resistivity with respect to its normal state value. The low-temperature resistivity under applied magnetic field is emphasized in the inset of Fig.4(a) and (b). As expected, $T_c$ shifts to lower temperature and the superconducting transition becomes broader with increasing magnetic field. Using the same criterion as for zero-field 
$\rho(T)$ data, we determined the temperature variation of the upper critical field ($\mu_0H_{c_2}$) for both compounds, presented in Fig.~\ref{fig:exp:resistivity}(c). The data were fitted with the Ginzburg-Landau expression \cite{SC_book2}: 
\begin{equation}
\mu_0H_{c_2}(T)=\mu_0H_{c_2}(0)\frac{1-t^2}{1+t^2}
\end{equation}
where $t = \frac{T}{T_c}$ and $T_c$ is the transition temperature at zero magnetic field. The relation fairly well describes the experimental data and one can obtain the values of $\mu_0H_{c_2}(0)$: $6.51(4)$\,T and $4.43(3)$\,T for SrIr$_2$ and SrRh$_2$, respectively.

Having the lower and upper critical fields, we find the coherence length with the formula $\xi_{GL}=\sqrt{\frac{\phi_0}{2\pi H_{c_2}}}$ and the penetration depth with self-consistent formula $\frac{4\pi H_{c_1}}{\phi_0}\lambda_{GL}^2=\ln\frac{\lambda_{GL}}{\xi_{GL}}$.
These parameters are equal to  $\xi_{GL}=71$\,\AA~and  $\lambda_{GL} = 890$\,\AA~in case of SrIr$_2$, while for SrRh$_2$ we have $\xi_{GL}=86$\,\AA~and $\lambda_{GL} = 1210$\,\AA. This gives the Ginzburg-Landau parameter $\kappa_{GL}=\frac{\lambda_{GL}}{\xi_{GL}}$ equal to $13$ for SrIr$_2$ and $14$ for SrRh$_2$ in agreement with the type-II nature of the studied superconductors.

All the above-described superconducting properties of studied materials are summarized in the Table  \ref{tab:exp}. For comparison, values obtained by Horie {\it et al.} and Gong {\it et al.} are also presented. 
  
 \begin{table}[b]
\caption{Lattice constant of Sr$M_2$ and $M$ ($M$ = Ir, Rh) obtained from experimental data $a_{\rm expt}$ and computed by optimization of the unit cell $a_{\rm calc}$, as well as the distances between the nearest atoms in the unit cell. All quantities are given in units of \AA.}
\label{tab:crystalinfo}
\begin{center}
\begin{ruledtabular}
\begin{tabular}{ l c c c c c }
& 
$a_{\rm expt}$ & 
$a_{\rm calc}$ & 
$M$-$M$ dist. & 
$M$-Sr dist.&
Sr-Sr dist.\\
\hline
{SrIr$_2$}& 7.793 & 7.753 &  2.775 & 3.254 & 3.399\\
{SrRh$_2$}&7.707  &7.781  &  2.721 & 3.190 & 3.332\\ 
{Ir \cite{ir-rh-lattice}}      &3.838  &3.883  & 2.714 & ---  & --- \\ 
{Rh \cite{ir-rh-lattice}}      &3.803  &3.866  &2.689  & ---  & --- \\ 
\end{tabular}
\end{ruledtabular}
\end{center}
\end{table}

\section{Electronic structure}
\subsection{Charge density and bonding}

\begin{figure}[t]
\includegraphics[width=0.5\textwidth]{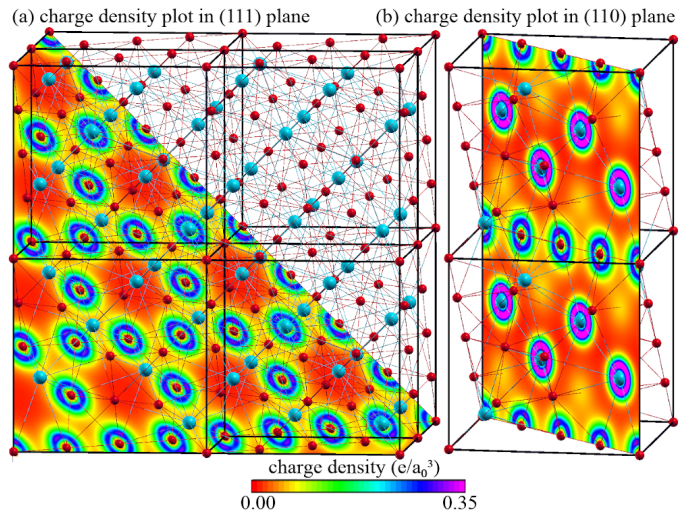}
\caption{Charge density (in $e/a_0^3$ units, where $e$ is the elementary charge and $a_0$ is the Bohr radius)} plotted in the supercell of SrIr$_2$ (a) on (111) plane, where the Ir hexagons are visible; and (b) on (110) plane, where the low-density space around Sr atoms is visible.\label{fig:calc:charge}
\end{figure}

In the first step of our theoretical study, the scalar-relativistic equilibrium unit cell volumes of both Sr$M_2$ compounds were calculated and obtained lattice constants are less than $1\%$ smaller than the experimental ones (see Table \ref{tab:crystalinfo}). Due to the larger size of Ir atom, the lattice constant of SrIr$_2$ is about $2\%$ larger than that of SrRh$_2$. As we have mentioned in the Introduction, the crystal structure of Sr$M_2$ is obtained from the 2x2x2 supercells of elemental {fcc} structures of $M$ by substitution of half of $M$ tetrahedrons with Sr atoms, placed in the center of mass of the removed tetrahedrons (see Fig.~\ref{fig:structure}). 
In Table \ref{tab:crystalinfo} the inter-atomic distances are compared. 
In Sr$M_2$, Ir (Rh) atoms have 6 nearest $M$ neighbors, distant by 2.78\,\AA \ (2.72\,\AA), in comparison to 12 atoms in the first coordination sphere in elemental { fcc} $M$ crystals, where the distances are slightly smaller (2.71 and 2.69\,\AA, respectively).
Similar distances and tetrahedron geometry suggest the similar character and strength of metallic $M$-$M$ bonding in both elemental metals and Laves phases.
The second coordination sphere of $M$ in Sr$M_2$ contains six Sr atoms distant by 3.25\,\AA \ (SrIr$_2$) and 3.19\,\AA \ (SrRh$_2$), while in the case of { fcc} $M$ crystals it contains 6 atoms distant by 3.84 and 3.80\,\AA, respectively. 
Sr in Laves phase is 12-fold coordinated by $M$ atoms (distances of 3.25 and 3.19\,\AA) and has 6 next-nearest Sr neighbors (distances of 3.40 and 3.33\,\AA).

\begin{figure*}[t]
\includegraphics[width=1.0\textwidth]{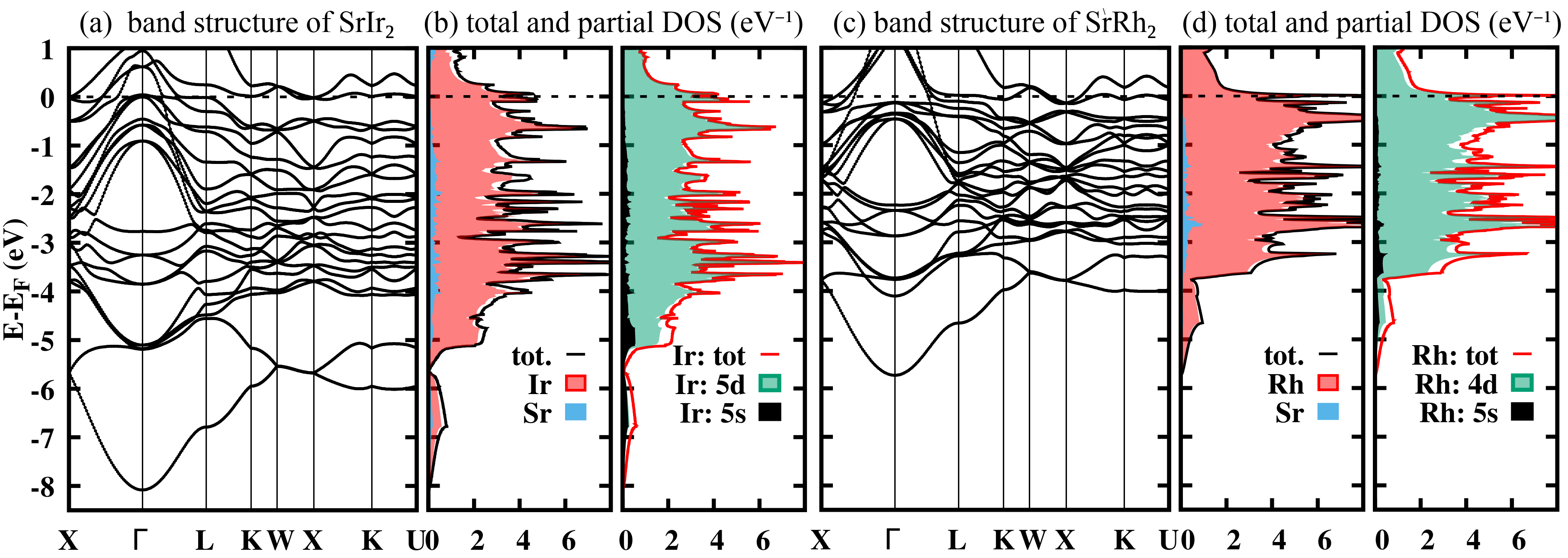}
\caption{Electronic structure of (a,b) SrIr$_2$;  and (c,d) SrRh$_2$ in terms of band
structure and densities of states. Total DOS is marked with black line and partial DOS is marked with colored lines. Results obtained with SOC.}\label{fig:calc:el}
\end{figure*}

Due to the substitution of a half of the dense-packed 4-atomic $M$ tetrahedrons in the elemental metallic structure by a single Sr alkaline earth atom, the structure becomes locally loosely-packed and a lot of ''free'' space around Sr atoms appears, with much lower charge density.
In Fig.~\ref{fig:calc:charge} valence charge density is plotted on (111) and (110) planes.
Originating from the elemental { fcc} structures of Ir or Rh, the hexagonal (111) metallic layers of $M$ atoms are seen in Fig.~\ref{fig:calc:charge}(a), which are separated by corrugated Sr-$M$ atomic layers (Fig.~\ref{fig:calc:charge}(b)). 
Here, each 12-fold coordinated Sr atom is located above or below the center of $M$ hexagon. As this is  a cubic structure, the hexagonal layers may be distinguished in all planes equivalent to (111) (all $M$ atoms are equivalent). 
The ''empty'' regions of low charge density around Sr are well seen in Fig.~\ref{fig:calc:charge}(b).

The Bader analysis, made using the {\sc Critic2} software \cite{critic}, shows the mixed ionic-covalent character of Ir-Sr bonding, as $1.27$ of two $6s$ electrons of Sr is transferred to Ir atoms, giving an additional $0.63$ electrons per Ir. This remains in agreement with the difference in the Pauling electronegativity of these elements ($2.20$ for Ir and $0.95$ for Sr). In spite of the different electronic configuration of free atoms ([Xe]$4f^{14}5d^76s^2$ for Ir and $[Kr]4d^85s^1$ for Rh) and smaller electron affinity of Rh ($1.14$\,eV) comparing to Ir ($1.57$\,eV) nearly the same charge transfer occurs for the case of SrRh$_2$ compound ($1.24$ of Sr electrons are transferred to two Rh atoms), leading to the same character of Rh-Sr bonding, and similar charge density distribution.

The impact of the described change of structure on the properties of the material is underlined by the value of bulk modulus $K_0$, calculated from Murnaghan equation of state \cite{murnaghan}, $P(V)=\frac{K_0}{K'_0}(\frac{V}{V_0}^{-K'_0}-1)$, where $V$ is a volume of the unit cell at pressure $P$, $V_0$ is volume at ambient pressure and $K'_0=\frac{\partial K_0}{\partial P}|_{P=0}$. The local reduction of the packing density of atoms by the formation of an "empty space" between the Ir tetrahedrons reduces the bulk modulus in the Laves phase from 324 GPa in Ir to 138 GPa in SrIr$_2$  (in a good agreement with the experimental value of $K_0=148.7$\,GPa~\cite{srir2-yang}), and from 244 GPa in Rh to 105 GPa in SrRh$_2$.

\subsection{Band structure of SrIr$_2$}

\begin{figure*}[t]
\includegraphics[width=1.00\textwidth]{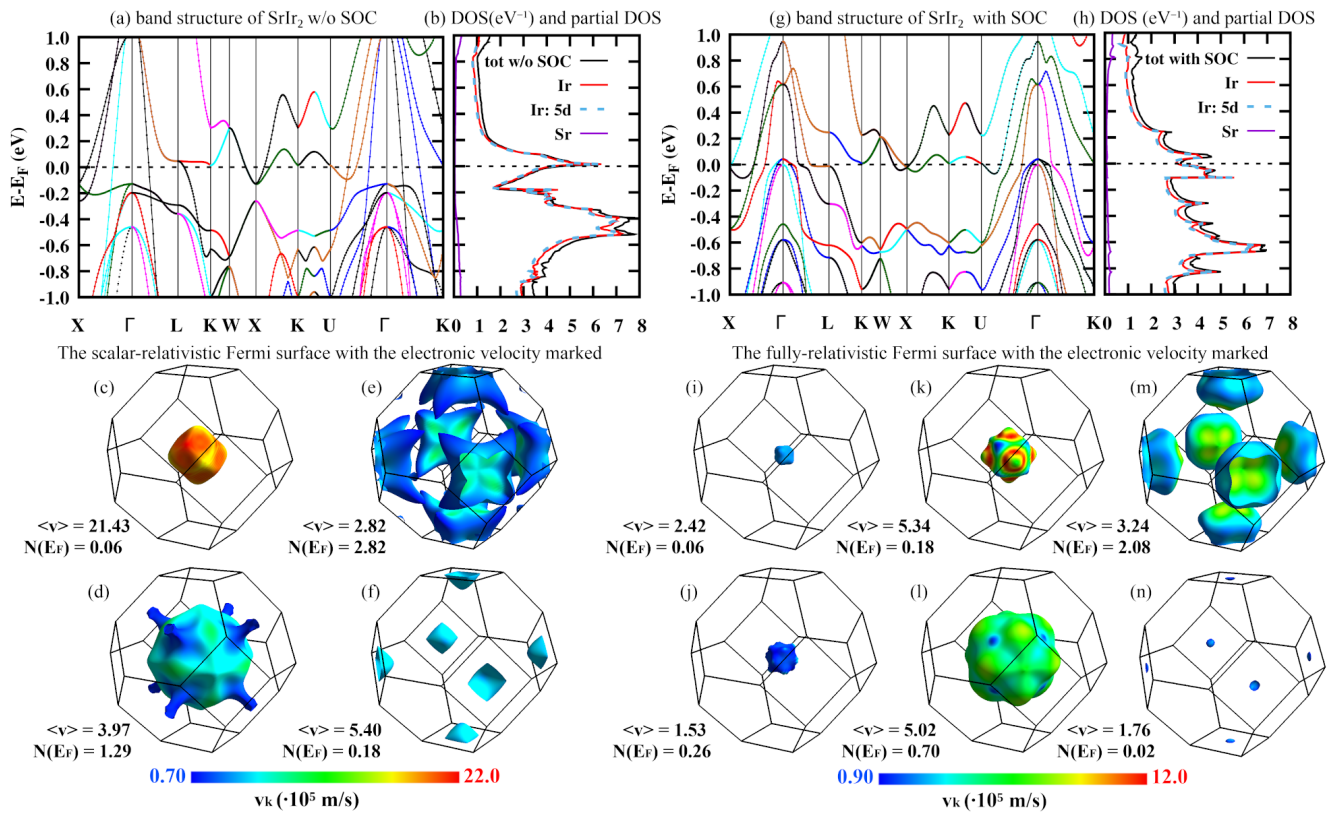}
\caption{The scalar-relativistic (a-f) and fully-relativistic (g-n) electronic structure of SrIr$_2$ in terms of
band structure (the points matched to each other with help of symmetry analysis are connected with
colored lines), DOS with atomic contributions and Fermi surface colored with respect to velocity. Average Fermi velocity $\langle v \rangle$ ($10^5$~m/s) and density of states $N(E_F)$ (eV$^{-1}$) for each of the FS sheets are shown.}\label{fig:calc:SrIr2-el}
\end{figure*}

Figures \ref{fig:calc:el} and \ref{fig:calc:SrIr2-el} present the calculated band structure, densities of states, and Fermi surface of SrIr$_2$. Densities of states are in fair agreement with those presented in Ref.~\cite{srir2-horie}. Sr atom contributes two $5s$ valence electrons, while Ir seven $5d$ electrons and two $6s$ electrons. In the crystal structure of SrIr$_2$, there are two formula units (f.u.) per primitive cell, thus there are $40$ valence electrons in the primitive cell, distributed among $\sim20$ occupied electronic bands. 
As shown in Fig.~\ref{fig:calc:el}(a), these bands span the energy range from $-8$\,eV to $E_F$, with a single band between -8 eV and -5.5 eV separated from the main valence band block. The partial atomic DOS shows that the valence band is contributed mostly by $d$ states of Ir, limiting the role of Sr to charge donor.

In Fig.~\ref{fig:calc:SrIr2-el} the large effect of SOC is shown by comparison of the band structure, DOS, and Fermi surface of SrIr$_2$ calculated without (a-f) and with SOC (g-n). 
In the scalar-relativistic case, the Fermi level is located on the slope of DOS and is crossed by four bands. This gives the four pieces of the Fermi surface shown in Fig.~\ref{fig:calc:SrIr2-el}(c-f), colored with respect to the Fermi velocity. 
There are two $\Gamma$-centered hole-like sheets [panels (c,d)] and two electron-like pieces with sheets centered at $X$ point [panels (e,f)]. 
When SOC is included [Fig.~\ref{fig:calc:SrIr2-el}(g-n)], the DOS peak, seen in the scalar-relativistic case, is split, and the Fermi level is located in between the two peaks in the local minimum of the DOS, which reduces $N(E_F)$ from $4.63$ eV$^{-1}$ per f.u. to $3.27$ eV$^{-1}$ per f.u. (see Table \ref{tab:dos}). 
Both the values and the reducing tendency after including SOC agree with DOS calculations of Ref.~\cite{srir2-horie}.
The Fermi surface of SrIr$_2$ now consists of six pockets in total (four $\Gamma$-centered hole-like pockets and two $X$-centered electron-like pockets). 
The additional two small, gamma-centered hole-like pockets (Fig.~\ref{fig:calc:SrIr2-el}(i-j)) not seen in the scalar-relativistic case appear due to a shift of the two low-velocity bands, which were placed just below $E_F$. 
The other four pieces of the Fermi surface (Fig.~\ref{fig:calc:SrIr2-el}(k-n)) are moderately modified when compared to their scalar-relativistic counter-parts in Fig.~\ref{fig:calc:SrIr2-el}(c-f). The partial DOS shown in Fig.~\ref{fig:calc:SrIr2-el}(b) confirms that Fermi surface is contributed by $d$ states of Ir.

From the obtained densities of states at the Fermi level, $N(E_F)$, the band structure values of the Sommerfeld coefficient \cite{grimvall} 
\begin{equation}
\gamma_{\rm band}=\frac{\pi^2}{3}{k_B^2}N(E_F)
\label{eq:lamgam0}
\end{equation}
are calculated and collected in Table \ref{tab:dos}. 
Comparing with the values obtained from the heat capacity measurements, $\gamma_{\rm expt}$, the electron-phonon coupling parameter $\lambda$ is estimated as \cite{grimvall} 
\begin{equation}
\gamma_{\rm expt} = \gamma_{\rm band}(1+\lambda_{\gamma}).
\label{eq:lamgam}
\end{equation}
In the case of SrIr$_2$, we obtain the value close to that expected based on the $T_c$ measurements, $\lambda_{\gamma} = 1.10$ only when SOC is included in the calculation of $N(E_F)$. This shows the importance of SOC for the analysis of SrIr$_2$, as the scalar-relativistic density of states would yield $\lambda_{\gamma} = 0.42$.

\begin{table}[b]
\caption{Electronic structure: the density of states at the Fermi level $N(E_F)$ (expressed in units of eV$^{-1}$ per f.u.), Sommerfeld coefficient: band structure value $\gamma_{\rm band}$ [Eq.~\ref{eq:lamgam0}] and the experimental value  $\gamma_{\rm expt}$ (both in $\frac{\text{mJ}}{\text{mol}\cdot\text{K}^2}$); Renormalization factor $\lambda_{\gamma}$ calculated from Sommerfeld coefficients using Eq.~\ref{eq:lamgam}.}
\label{tab:dos}
\begin{center}
\begin{ruledtabular}
\begin{tabular}{ l r c c c c c }
&\multicolumn{2}{c}{w/o SOC}&\multicolumn{2}{c}{with SOC}&\multicolumn{2}{c}{with SOC}\\
 &{SrIr$_2$} & {SrRh$_2$} & {SrIr$_2$} &
{SrRh$_2$} & Ir & Rh\\
\hline
$N(E_F)$ &4.63 &  5.52 &3.27 & 5.58 & 0.93 & 1.42\\
$\gamma_{\rm band}$ &10.92 & 13.02 &7.72 & 13.14 & 2.20 & 3.35\\
$\gamma_{\rm expt}$ & &  &15.50 & 27.30 & 3.27 \footnote{\label{ir-rh}\citet*{heat-metals}} & 4.65\footref{ir-rh}\\
$\lambda_{\gamma}$ & 0.42 & 1.10 &1.01 & 1.08 & 0.49 & 0.39\\
\end{tabular}
\end{ruledtabular}
\end{center}
\end{table}

\subsection{Band structure of SrRh$_2$}

 \begin{figure*}[t]
\includegraphics[width=1.00\textwidth]{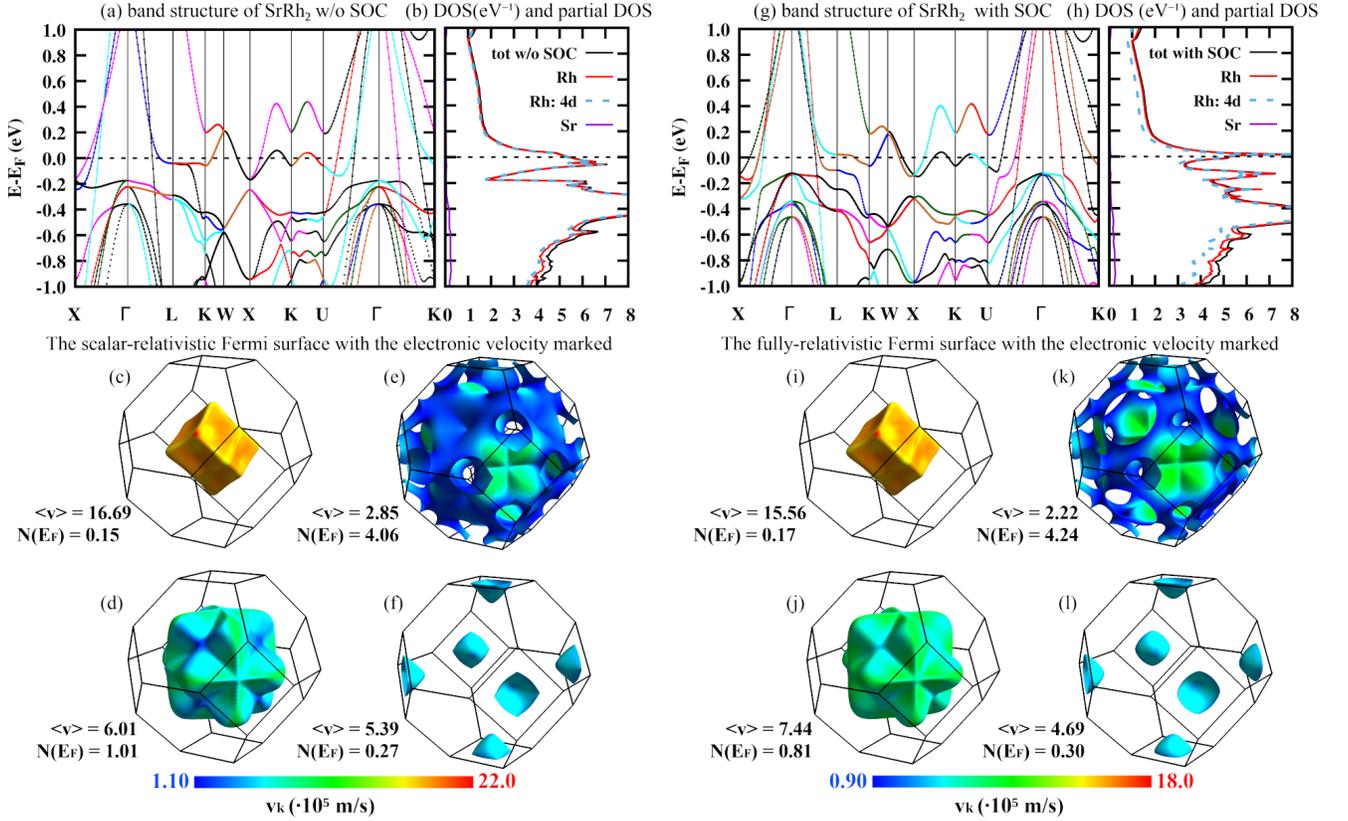}
\caption{The scalar-relativistic (a-f) and fully-relativistic (g-l) electronic structure of SrRh$_2$ in terms of
band structure (the points matched to each other with help of symmetry analysis are connected with
colored lines), DOS with atomic contributions and Fermi surface colored with respect to velocity. Average Fermi velocity $\langle v \rangle$ ($10^5$~m/s) and density of states $N(E_F)$ (eV$^{-1}$) for each of the FS sheets are shown.}\label{fig:calc:SrRh2-el}
\end{figure*}

Band structure and DOS of SrRh$_2$ are plotted in Fig.~\ref{fig:calc:el} next to that of SrIr$_2$ for a convenient comparison, whereas more details near the Fermi energy are shown in Fig.~\ref{fig:calc:SrRh2-el}.
Because of their isoelectronic nature, both Laves phases have qualitatively similar electronic structures. However, in correlation with the presence of $4d$ orbitals in Rh, instead of $5d$ in Ir, valence bands of SrRh$_2$ are less dispersive and spread over a smaller energy range of about $6$\,eV. Two electrons occupy the lowest band, which here extends from $-6$\,eV to $-3.8$\,eV with respect to the Fermi level [Fig.~\ref{fig:calc:el}(c)]. 
The remaining valence band block, accommodating $38$ electrons, spans the relatively small energy range from $-3.8$\,eV to $E_F$.
This results in generally higher DOS values, when compared to Ir analog which has a larger bandwidth: $N(E_F) = 5.58$ eV$^{-1}$ in SrRh$_2$ versus $3.27$ eV$^{-1}$ in SrIr$_2$ (see Table~\ref{tab:dos}).
The scalar-relativistic band structure of SrRh$_2$ is very similar to that in SrIr$_2$ around the Fermi level, and only the DOS peak appears slightly below $E_F$, as seen in Fig.~\ref{fig:calc:SrRh2-el}(a).
More differences appear for the fully-relativistic case. As rhodium is much lighter than iridium, the electronic properties of SrRh$_2$ are weakly affected by SOC, in contrast to the case of SrIr$_2$ discussed above. 
When SOC is included, the DOS peak becomes narrower but the Fermi level remains on the slope of the peak in both scalar- and fully-relativistic case. The value of $N(E_F)$ is almost the same in both cases, $5.52$ and $5.58$ eV$^{-1}$ per f.u. respectively.
As in the case of SrIr$_2$, the scalar-relativistic Fermi surface consists of four pieces, and this number remains also in the relativistic case, with FS hardly affected by SOC.  
The obtained value of $N(E_F)$ yields the electron-phonon coupling constant equal to $\lambda_{\gamma}=1.08$ when computed as a renormalization factor of the Sommerfeld parameter (see Table~\ref{tab:dos}).

\subsection{Relation to elemental Ir and Rh}
Following the close relation of the crystal structures, we find similarities in the electronic structures of Sr$M_2$ and elemental metallic $M = $ Ir, Rh, which we present in Appendix, Fig.~\ref{fig:ir-rh-el}.
There we may also distinguish the lowest, separated band and a characteristic spiky DOS structure. 
The occupied bands of Ir and Rh spread the wider energy range of $11.5$\,eV (Ir) and $8$\,eV (Rh), due to their closed-packed crystal structure and more metallic character.
When we consider the zoom of DOS of Sr$M_2$ shown in Figs.~\ref{fig:calc:SrIr2-el} -- \ref{fig:calc:SrRh2-el} and compare with the DOS of Ir and Rh shown in Fig.~\ref{fig:ir-rh-el}, the characteristic DOS peak near the Fermi level is present in all cases. In the elemental structures of Ir and Rh, $E_F$ is below the peak, whereas in Sr$M_2$ addition of Sr acts as an electron doping, pushing $E_F$ into the peak and increasing the DOS per $M$ atom, from 0.93\,eV$^{-1}$ (1.42\,eV$^{-1}$) for Ir (Rh) to 1.63\,eV$^{-1}$ (2.79\,eV$^{-1}$) for SrIr$_2$ (SrRh$_2$), see Table \ref{tab:dos}.
Even though the Fermi level is pushed, similarities in the shape of the Fermi surface may be also noted, especially when comparing SrRh$_2$ to Rh.
Concluding, the band structures of Sr$M_2$ near $E_F$ have a lot in common with the elemental metallic $M$ structures and Sr atom plays a role of an electron donor.

\section{Phonons and electron-phonon coupling}

\subsection{SrIr$_2$}

\begin{figure*}[t]
\includegraphics[width=0.80\textwidth]{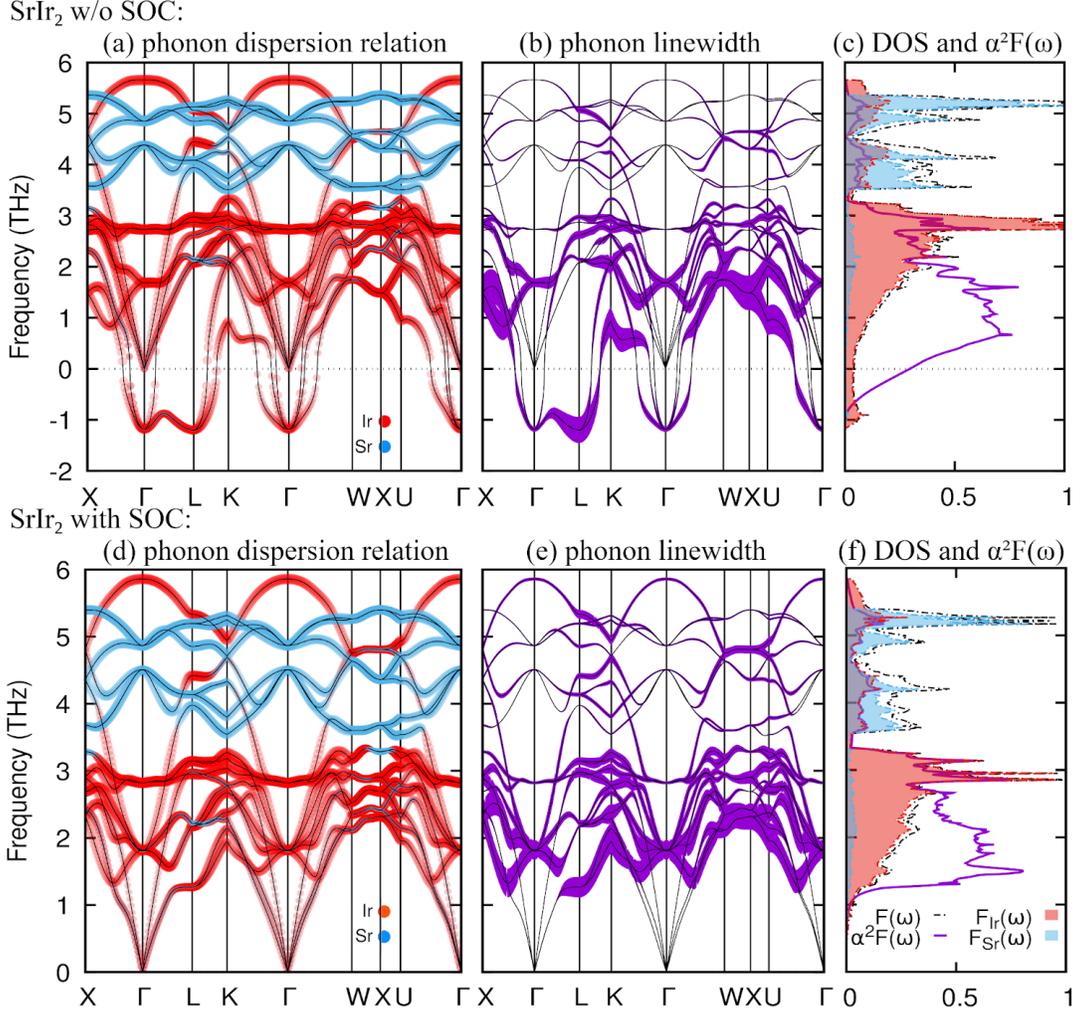}
\caption{The phonon structure and EPC of SrIr$_2$ determined (a-c) without SOC;
and (d-f) with SOC. Panels (a, c): phonon dispersion relations with contributions from Sr (in blue) and Ir (in red); panels (b,e) phonon dispersion relations with band shading 
proportional to the phonon linewidths ($\gamma_{{\bf q}\nu}$ in THz multiplied by 4); panels (c,f) total and partial phonon DOS, $F(\omega)$, and Eliashberg function, $\alpha^2F(\omega)$. 
 To conveniently visualize both $F(\omega)$ (which originally is expressed in units of THz$^{-1}$ and has a norm of $3N_{at}$, where $N_{at}$ - number of atoms in the primitive cell)  and $\alpha^2F(\omega)$ (which is dimensionless) in one panel, the former was renormalized as  $\alpha^2F(\omega)$. }\label{fig:calc:SrIr2-elph}
\end{figure*}

The phonon dispersion relations and phonon DOS $F(\omega)$ of SrIr$_2$ are shown in Fig.~\ref{fig:calc:SrIr2-elph}. 
The phonon spectrum of SrIr$_2$ is not stable in the scalar-relativistic case, as in Fig.~\ref{fig:calc:SrIr2-elph}(a) we observe imaginary frequencies (here plotted as negative) around 
$\Gamma = (0, 0, 0)$ and $L = (0.5, 0.5, 0.5)$ $\bf{q}$-points, associated with the three Ir optical modes. 
The structure is then stabilized when SOC is included [Fig.~\ref{fig:calc:SrIr2-elph}(d)]. 
The phonon spectrum spans the range from 0 to nearly 6 THz and is composed of 18 modes (there are six atoms in the primitive cell).
The lower- and higher-frequency parts are separated by a pseudo-gap, formed around 3.4 THz in the phonon density of states in Fig.~\ref{fig:calc:SrIr2-elph}(f). 
Modes below 3.4 THz originate mostly from Ir, while those between 3.4 THz and 5.5 THz are related mostly to Sr vibrations, reflecting the differences in their atomic masses ($M_{\rm Ir} = 192.2$~u, $M_{\rm Sr} = 87.6$~u).
In contrast, the highest-frequency optical mode, which extends from 4.5 THz to 6 THz, again involves the vibrations of the heavier Ir.
This characteristic ''bell-shaped'' mode, as we discuss below, is also present in the elemental { fcc} Ir, and its observation guided us to take a closer look at the similarities between SrIr$_2$ and Ir.

To characterize the phonon spectrum, several phonon frequency moments are calculated 
using the following formulas:
\begin{equation}\label{eq:mom}
\langle \omega^n \rangle = \int_0^{\omega_{\mathsf{max}}} \omega^{n-1} F(\omega) d\omega \left/ \int_0^{\omega_{\mathsf{max}}} F(\omega) \frac{d\omega}{{\omega}} \right.,
\end{equation}

\begin{equation}\label{eq:sred}
\langle \omega \rangle = \int_0^{\omega_{\mathsf{max}}} \omega F(\omega) d\omega \left/ \int_0^{\omega_{\mathsf{max}}} F(\omega) d\omega \right.,
\end{equation}

\begin{equation}\label{eq:omlog}
\langle\omega_{\rm log}\rangle = \exp\left(\int_0^{\omega_{\mathsf{max}}} F(\omega) \ln\omega\frac{d\omega}{{\omega}} \left/ \int_0^{\omega_{\mathsf{max}}} 
{F(\omega)}\frac{d\omega}{{\omega}} \right. \right),
\end{equation}

and

\begin{equation}\label{eq:omlog2}
\langle\omega_{\rm log}^{\alpha^2F}\rangle = \exp\left(\int_0^{\omega_{\mathsf{max}}} \alpha^2F(\omega) \ln\omega\frac{d\omega}{{\omega}} \left/ \int_0^{\omega_{\mathsf{max}}} 
{\alpha^2F(\omega)}\frac{d\omega}{{\omega}} \right. \right).
\end{equation}
Results are shown in Table \ref{tab:freq}. The global average phonon frequency is 3.38 THz, whereas the partial for Ir and Sr is 2.96 THz and 4.22 THz, respectively.

\begin{table}[b]
\caption{The phonon frequency moments of SrIr$_2$, SrRh$_2$, Ir and Rh, obtained with help of Eq.(\ref{eq:mom})--(\ref{eq:omlog2}).}
\label{tab:freq}
\begin{center}
\begin{ruledtabular}
\begin{tabular}{ c c c c c c }
& 
$\langle\omega^1\rangle$ & 
$\sqrt{\langle\omega^2\rangle}$ & 
$\langle\omega\rangle$ & 
$\langle\omega_{\rm log}\rangle$ & 
$\langle\omega_{\rm log}^{\alpha^2F}\rangle$ \\
\hline
\multicolumn{6}{c}{SrIr$_2$ w/o SOC (THz)}\\ 
\multicolumn{6}{c}{unstable (imaginary frequencies)}\\
\multicolumn{6}{c}{SrIr$_2$ with SOC (THz)}\\ 
total         & 2.88 & 3.12 & 3.38 & 2.62 & 2.02 \\
Sr         & 3.78 & 4.00 & 4.22 & 3.47 &  \\
Ir         & 2.57 & 2.76 & 2.96 & 2.38 &  \\
\hline
\multicolumn{6}{c}{SrRh$_2$ w/o SOC (THz)}\\ 
total         & 2.98 & 3.21 & 3.44 & 2.74 & 2.24 \\
\multicolumn{6}{c}{SrRh$_2$ with SOC (THz)}\\ 
total         & 3.08 & 3.29 & 3.51 & 2.85 & 2.52 \\
Sr         & 3.28 & 3.50 & 3.74 & 3.00 &  \\
Rh         & 3.00 & 3.19 & 3.40 & 2.78 &  \\
\hline
\multicolumn{6}{c}{Ir with SOC (THz)}\\ 
total         & 4.36 & 4.5 & 4.65 & 4.17 & 4.61 \\
\hline
\multicolumn{6}{c}{Rh with SOC (THz)}\\ 
total         & 4.83 & 5.0 & 5.17 & 4.61 & 5.04    \\
\end{tabular}
\end{ruledtabular}
\end{center}
\end{table}

\begin{figure*}[t]
\includegraphics[width=0.99\textwidth]{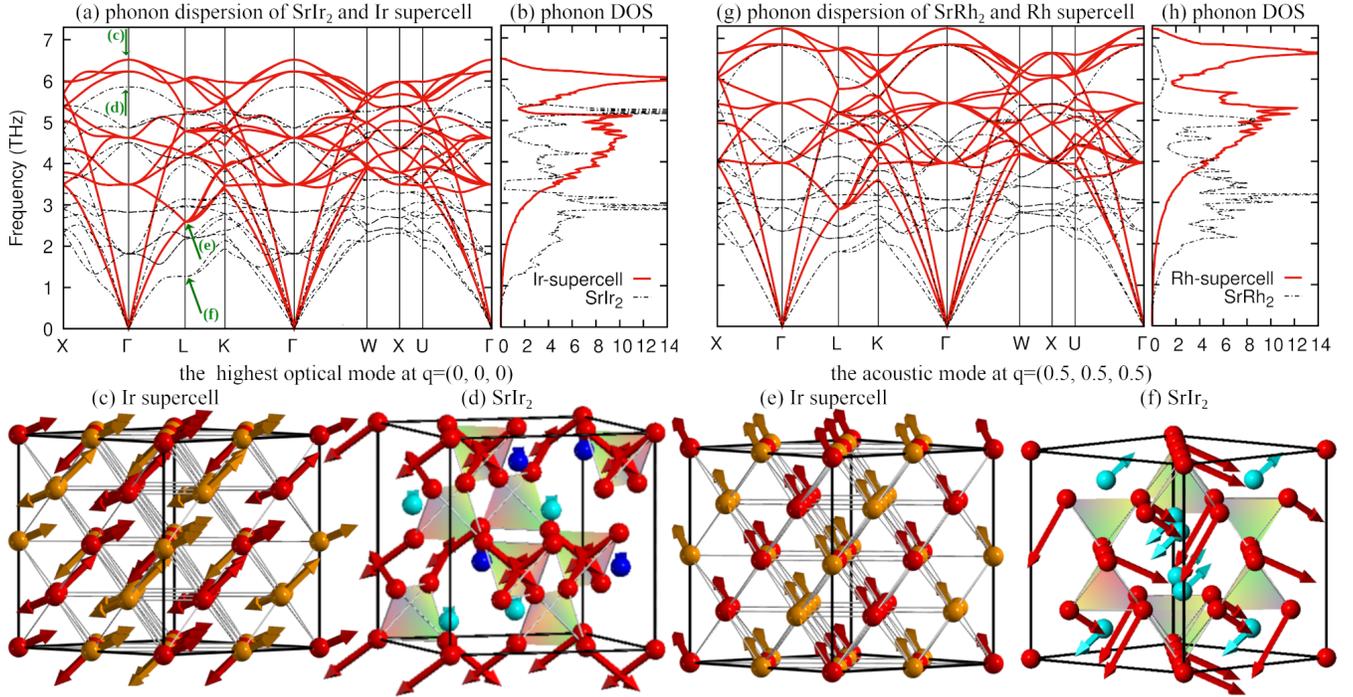}
\caption{The phonon dispersion relation and DOS (a-b) of both {fcc} Ir 2x2x2 supercell and  SrIr$_2$. Additionally the phonon modes  at $\Gamma=(0,0,0)$ and L$=(0.5, 0.5, 0.5)$, marked in (a) with green arrows, are  shown in real space in supercell of Ir (c, e) and in unit cell of SrIr$_2$ (d, f) with tetrahedra of atoms  marked. The phonon dispersion relation and phonon DOS of Rh supercell and of SrRh$_2$ are shown in panels (g-h).
}\label{fig:calc:ir-srir2-elph}
\end{figure*}

The electron-phonon coupling matrix elements $g_{{\bf q}\nu}({\bf k},i,j)$ are next calculated as \cite{wierzbowska,heid-pb,gustino-rmp}
\begin{equation}\begin{split}
g_{{\bf q}\nu}({\bf k},i,j)& =\\
=\sum_s &\sqrt{{\hbar\over 2M_s\omega_{{\bf q}\nu}}}
\langle\psi_{i,{\bf k+q}}| {dV_{\rm SCF}\over d {\hat u}_{\nu s} }\cdot
                   \hat \epsilon_{\nu s}|\psi_{j,{\bf k}}\rangle,
\label{calc:el-ph-matrix}
\end{split}\end{equation}
where  $i,j$ are band indexes, $M_s$ is a mass of atom $s$, $ {dV_{\rm SCF}\over d {\hat u}_{\nu s} }$ is a change of electronic potential calculated in self-consistent cycle due to the movement of an atom $s$, $\hat{\epsilon}_{\nu s}$ is a polarization vector associated with $\nu$-th phonon mode ${\hat u}_{\nu s}$ and $\psi_{i,{\bf k}}$ is the electronic wave function.
On this basis the phonon linewidths $\gamma_{{\bf q}\nu}$ are calculated by summing $g_{{\bf q}\nu}({\bf k},i,j)$ over all the electronic states on the Fermi surface, which may interact with the given phonon $\{{\bf q}\nu\}$ \cite{wierzbowska,grimvall,gustino-rmp}:
\begin{equation}
\begin{split}
\gamma_{{\bf q}\nu} =& 2\pi\omega_{{\bf q}\nu} \sum_{ij}
                \int {d^3k\over \Omega_{\rm BZ}}  |g_{{\bf q}\nu}({\bf k},i,j)|^2 \\
                    &\times\delta(E_{{\bf k},i} - E_F)  \delta(E_{{\bf k+q},j} - E_F).
\end{split}
\end{equation}

Phonon linewidths are visualized in Fig.~\ref{fig:calc:SrIr2-elph}(b,e). 
As we can see, the strongest electron-phonon interactions (largest $\gamma_{{\bf q}\nu}$) are observed for the lowest acoustic and lowest optical Ir modes, which we discuss in more detail in the next Section. 
None of the Sr-dominated modes show large $\gamma_{{\bf q}\nu}$.

 \begin{figure*}[t]
\includegraphics[width=0.99\textwidth]{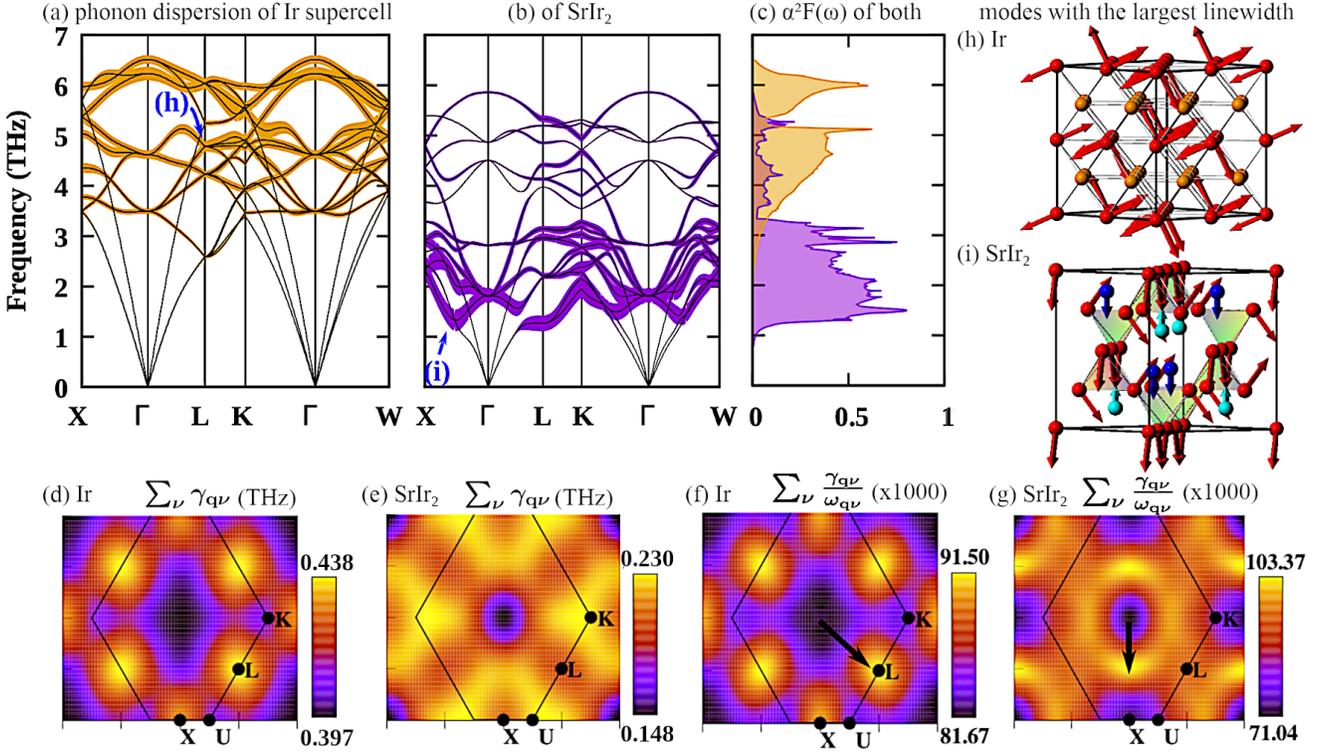}
\caption{The connection of phonon linewidths with real-space vibrations of atoms. Phonon dispersions with shading proportional to the phonon linewidths ($\gamma_{{\bf q}\nu}$ in THz multiplied by 4) in (a) 2x2x2 supercell of Ir; (b) and SrIr$_2$; (c) Eliashberg function of both crystals. Panels (d) and (e) show maps in (1, $\overline{1}$, 0) plane of BZ of phonon linewidth summed over modes $\sum_{\nu}\gamma_{{\bf q}\nu}$ of Ir and SrIr$_2$, respectively, while panels (f) and (g) show maps of $\sum_{\nu}\frac{\gamma_{{\bf q}\nu}}{\omega_{{\bf q}\nu}}$; (h) and (i) show visualization of modes with the largest linewidths: (h) mode at q = (0.5, 0.5, 0.5) of crystalline Ir; and (i) mode at q = (0.5, 0.0, 0.0) of SrIr2. Both are marked in phonon dispersion relations (a) and (b).}\label{fig:ir-elph}
\end{figure*}

In the next step, the Eliashberg function $\alpha^2F(\omega)$ is calculated as a sum of $\gamma_{{\bf q}\nu}$ over all phonon modes, weighted by the inverse of their frequency \cite{wierzbowska,grimvall,gustino-rmp}:
\begin{equation}
\alpha^2F(\omega) = {1\over 2\pi N(E_F)}\sum_{{\bf q}\nu} 
                    \delta(\omega-\omega_{{\bf q}\nu})
                    {\gamma_{{\bf q}\nu}\over\hbar\omega_{{\bf q}\nu}}.
\end{equation}
Eliashberg function $\alpha^2F(\omega)$ is plotted together with the phonon DOS $F(\omega)$ in 
Fig.~\ref{fig:calc:SrIr2-elph}(c,f). 
Due to the large $\gamma_{{\bf q}\nu}$ and low frequencies, in Fig.~\ref{fig:calc:SrIr2-elph}(f) the largest $\alpha^2F(\omega)$, much enhanced above $F(\omega)$, is seen in the Ir-dominated part from 0 to 2.8 THz. 
At higher frequencies  $\alpha^2F(\omega)$ follows the shape of the partial Ir DOS function, $F_{Ir}(\omega)$, as in the case of crystalline Ir, discussed below.

Finally, the EPC constant $\lambda$ is calculated as the integral of the Eliashberg function divided by frequency \cite{grimvall},
\begin{equation}\label{eq:lam2}
\lambda=2\int_0^{\omega_{\rm max}} \frac{\alpha^2F(\omega)}{\omega} \text{d}\omega.
\end{equation}
The obtained $\lambda = 1.09$ indicates that the electron-phonon coupling is strong in this compound, and is in very good agreement with the value of 1.17, calculated from experimental $T_c$ with help of the Allen-Dynes equation, alike with $\lambda_{\gamma} = 1.01$, extracted from renormalization of the Sommerfeld coefficient. These values are presented in Tables \ref{tab:dos} and \ref{tab:tc}.

\subsection{Phonon engineering}

As we presented in Fig.~\ref{fig:structure}, the structure of SrIr$_2$ is formed as a 2x2x2 supercell of Ir in which half of the Ir tetrahedrons are replaced by Sr atoms.
It is then intriguing to see how this process modifies the lattice dynamics of Ir, transforming the poorly superconducting material ($T_c = 0.14$~K) into the strongly-coupled superconductor. 

To do this, we have calculated the phonon structure and electron-phonon coupling in elemental { fcc} Ir (as well as Rh, discussed later).
Phonon dispersion relations with the phonon linewidths, DOS, and Eliashberg function for Ir are shown in Fig.~\ref{fig:ir-rh-elph} in Appendix. 
However, as the direct comparison of phonon dispersion relations between the monoatomic Ir and multiatomic SrIr$_2$ would be impossible, 
we have additionally recalculated the phonon dispersion curves for the Ir 2x2x2 supercell, which now may be plotted together with those of SrIr$_2$ in Fig.~\ref{fig:calc:ir-srir2-elph}~\footnote{Note, that phonon dispersion relations of Ir and Rh in Figs.~\ref{fig:calc:ir-srir2-elph}, \ref{fig:ir-elph} are shown in the Brillouin zones of 8-atom 2x2x2 supercells formed from primitive cell,  whereas the phonon modes are visualized in the conventional cubic 2x2x2 supercells, containing 32 atoms. For the relation between those two representations see Fig. S2 in Supplemental Material~\cite{supplemental}}.
The electron-phonon couplings of the two structures are compared in Fig.~\ref{fig:ir-elph}.
The phonon structure of elemental Ir consists of three acoustic modes in the frequency range from 0 to about 6.5 THz, i.e. nearly the same range as the whole spectrum of SrIr$_2$. 
When the phonon dispersion curves of Ir are folded into the Brillouin zone (BZ) of the supercell in Fig.~\ref{fig:calc:ir-srir2-elph}, we see that in the higher frequency range ($\omega > 4$ THz) several phonon branches are recreated in SrIr$_2$ after the Ir$_4 \rightarrow$ Sr substitution. Due to the reduced population of atoms, the number of phonon branches is smaller in the Laves phase, but their dispersions are similar to those in Ir, with the frequencies modified up to $\sim 10\%$. 
This includes the highest ''bell-shaped'' mode with the maximum frequency at $\Gamma$ point. 
In the case of SrIr$_2$ [see Fig.~\ref{fig:calc:ir-srir2-elph}(d)] this mode is related to the displacements of Ir atoms towards the center of the closed-packed Ir tetrahedron, which explains its high frequency. 
Similar mode was observed in CaPd$_2$ Laves phase in Ref.~\cite{capd2}, where it was discussed in the context of a large chemical pressure induced by such atomic vibrations.
Strong deformations of the close-packed Ir tetrahedron are associated with the highest-frequency modes in the Ir structure, one of which is visualized in Fig.~\ref{fig:calc:ir-srir2-elph}(c) and the others in Fig. S3 in Supplemental Material \cite{supplemental}.

\begin{figure}[t]
\includegraphics[width=0.45\textwidth]{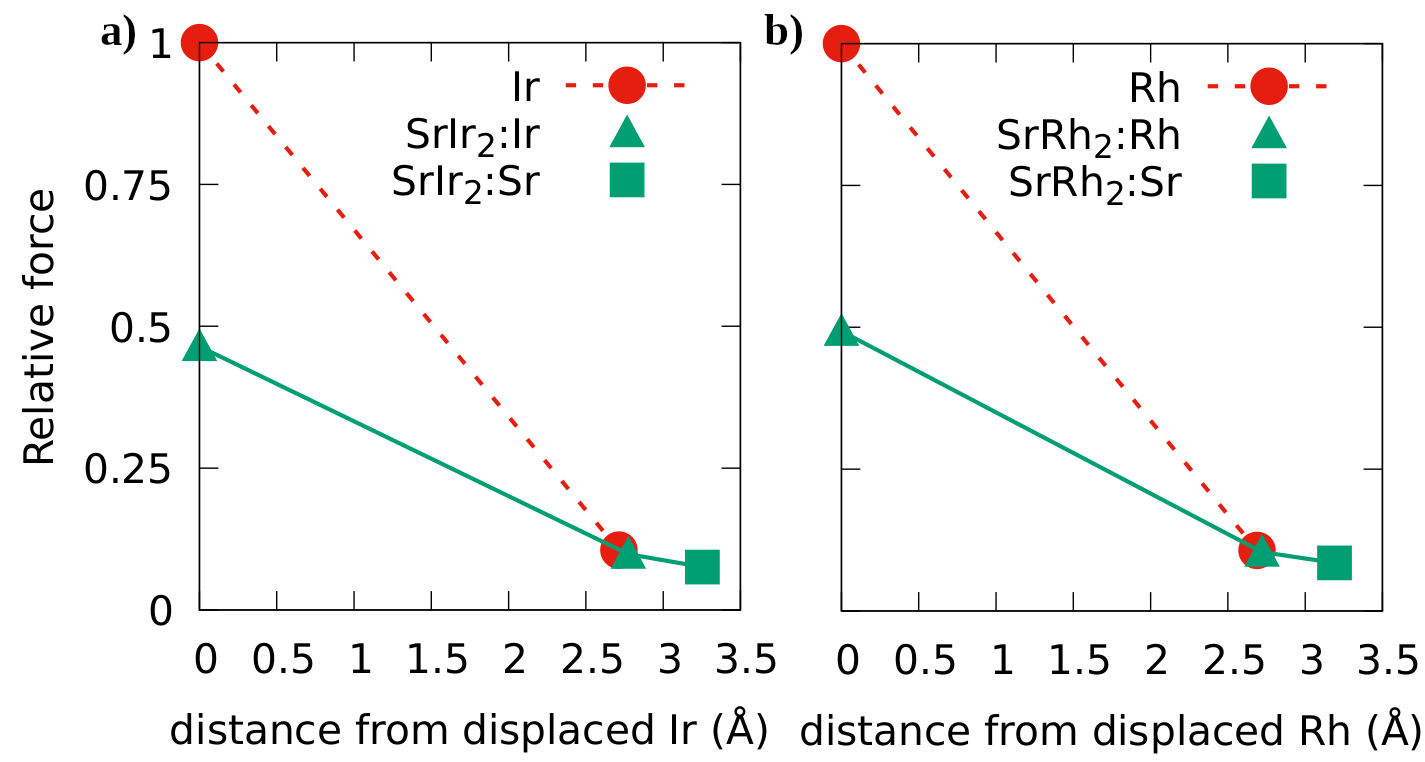}
\caption{(a) The forces acting on the nearest atoms in the unit cell when Ir is displaced, normalized to the force in elemental metallic Ir. The restoring force, acting on the displaced atom, is about two times smaller in SrIr$_2$, compared to Ir, whereas the force acting on neighbors (marked on the $x$ axis according to their distance from the displaced atom) is comparable;
(b) The same for { fcc} Rh and SrRh$_2$ when Rh is displaced. }\label{fig:force-constant}
\end{figure}

In the lower frequency range, on the other hand, the phonon dispersion curves in SrIr$_2$ are completely changed as they have about $50\%$ lowered frequencies when compared to the crystalline Ir. This effect is directly related to reduction of the number of Ir-Ir neighbors and a much less-packed crystal structure around the substituted Sr atoms. More space around the Ir tetrahedrons locally reduces the internal chemical pressure connected to atomic vibrations.
For example, we can compare the two lowest acoustic modes at L-point, indicated by arrows in Fig.~\ref{fig:calc:ir-srir2-elph}(a) and visualized in real space in Fig~\ref{fig:calc:ir-srir2-elph}(e,f).
In the Ir supercell, this mode is associated with vibrations along the [2, 1, 1] direction, in which every second atom in the Ir chains vibrate perpendicular to the chain.
Such a mode is less energetic than that involving larger tetrahedron deformation, but still due to 12-fold coordination of Ir atoms, the forces acting on the displaced atoms inside such chains are strong, resulting in the frequency of 2.6 THz at L.
In SrIr$_2$, on the other hand [Fig.~\ref{fig:calc:ir-srir2-elph}(f)], in the lowest mode at L-point
Ir atoms vibrate towards the ''empty space'' with a low charge density, formed when Ir$_4$ tetrahedron was substituted by Sr. This does not involve such large energy, thus the frequency is reduced to 1.3 THz.
Moreover, in the Laves phase each Ir has only 6 nearest Ir neighbors, thus even though the individual Ir-Ir bond strength is similar in both cases of metallic iridium and the Laves phase, the restoring force acting on atoms in the case of a simple displacement of a single atom is much reduced. 
This is shown in Fig.~\ref{fig:force-constant}.
Single Ir atom is displaced and the restoring force in SrIr$_2$ is about two times smaller than in crystalline Ir (the force in Ir was taken as unity).
Importantly, the forces acting on the nearest Ir neighbors are similar in both cases, confirming the same bond strength between Ir atoms which form tetrahedrons, while the Ir-Sr bonding occurs to be weaker.

As a consequence of the reduced coordination and density of atom packing, the average phonon frequency of Ir in SrIr$_2$ is lowered compared to metallic Ir (3.38 THz versus 4.65 THz, see Table \ref{tab:freq}).  
Thus, the replacement of half of Ir tetrahedrons with Sr atoms is the key factor 
determining the phonon structure of SrIr$_2$.

\begin{table}[b]
\caption{Superconducting properties of SrIr$_2$, SrRh$_2$, Ir, and Rh, in terms of the EPC constant $\lambda$ and critical temperature $T_c$ determined theoretically [Eq. (\ref{eq:tc}) and Eq. (\ref{eq:lam2})] and experimentally. In each case $\mu^*=0.13$ was assumed.
For the Laves phases the experimental value of $\lambda$ is calculated from the measured $T_c$ using Eq. (\ref{eq:tc}) and $\langle\omega_{\text{log}}^{{\alpha^2F(\omega)}}\rangle$ from Eq. (\ref{eq:marsiglio}). For Ir and Rh $\lambda$ was determined in Ref~\cite{rhodium-superconductivity}  based on McMillan's Eq. (\ref{eq:mcmillan}) and much lower $T_c$ of Rh is due to presence of spin fluctuations with $\lambda_{\rm sf} \sim 0.1$.}
\label{tab:tc}
\begin{center}
\begin{ruledtabular}
\begin{tabular}{l c c c c }
& \multicolumn{2}{c}{calculated}&\multicolumn{2}{c}{experimental}\\
& 
$\lambda$ & 
$T_{c}$ (K) & 
$\lambda$ &
$T_{c}$ (K)  \\
\hline
{SrIr$_2$ w/o SOC}&\multicolumn{2}{c}{\centering unstable}\\ 
{SrIr$_2$ w. SOC}&1.09&6.88&1.17&6.07\\ 
{SrRh$_2$ w/o SOC}&1.12&8.00\\ 
{SrRh$_2$ w. SOC}&0.90&5.93&0.93&5.41\\ 
{Ir w. SOC}&0.36&0.17&0.34&0.14\\ 
{Rh w. SOC}&0.36&0.19&0.34&0.3$\times 10^{-3}$
\end{tabular}
\end{ruledtabular}
\end{center}
\end{table}

Modifications of the phonon spectrum lead to strong changes in the electron-phonon interactions.
The Eliashberg function of Ir closely follows its phonon DOS [see Fig. \ref{fig:ir-rh-elph}(b)], nearly satisfying the relation $\alpha^2F(\omega) \simeq {\rm const} \times F(\omega)$. 
In the Laves phase, such behavior is observed only in the high-frequency part, whereas at lower $\omega$ the electron-phonon interaction is more frequency-dependent with $\alpha^2F(\omega)$ strongly enhanced above $F(\omega)$ (see Fig. \ref{fig:calc:SrIr2-elph}). This difference is clearly visible in Fig. \ref{fig:ir-elph}, where the Eliashberg functions are compared. 
As $\alpha^2F(\omega)$ depends on the ratio $\frac{\gamma_{{\bf q}\nu}}{\omega_{{\bf q}\nu}}$,
the significant decrease in phonon frequencies of Ir in SrIr$_2$ may enhance the EPC 
under the assumption that the phonon linewidths, which are essentially an electronic contribution to the electron-phonon coupling, are not much affected. 
Phonon linewidths marked as shading in the phonon dispersion relation (panels a-b) as well as summed over all modes (d-e) show, that the phonon linewidths are generally larger in the case of crystalline Ir (compare the scale of panels d and e). Here, all optical modes of Ir have large $\gamma_{{\bf q}\nu}$, 
whereas in the case of SrIr$_2$ only low frequency modes count. 

To compare more global values and analyze how the electronic contribution changes upon Sr insertion to the structure, we calculate the first moment of the Eliashberg function as the integral $I$, which is proportional to the sum of phonon linewidths over both {\bf q} and $\omega$:
\footnote{This integral is closely related to the so-called McMillan-Hopfield parameter $\eta$~\cite{mcmillan,grimvall}, which for a monoatomic system is defined as $\eta=2MI$, with $M$ being the atomic mass and $I$ defined as in Eq.(\ref{eq:I}). EPC constant is then represented using the well-known formula $\lambda = \eta/M\langle\omega^2\rangle$, where $\langle\omega^2\rangle$ in a system where EPC is frequency-independent may be calculated using Eq.(\ref{eq:mom}) with $n=2$.}

\begin{equation}
I=\int_0^{\omega_{\rm max}} \omega\cdot\alpha^2F(\omega){\rm d}\omega.
\label{eq:I}
\end{equation}
$I$ is the quantity which does not depend on phonon frequency, as 

\begin{equation}
\begin{split}
I=&\frac{1}{2\pi\hbar N(E_F)} \int_0^{\omega_{max}} d\omega
                    \sum_{{\bf q}\nu}
                    \delta(\omega-\omega_{{\bf q}\nu}) \gamma_{\textbf{q}\nu}=\\
=& \frac{1}{2\pi\hbar N(E_F)} \int_0^{\omega_{max}} d\omega
                    \sum_{{\bf q}\nu}
                    \delta(\omega-\omega_{{\bf q}\nu})
                     \sum_s \frac{1}{2M_s} \times \\
&\times\int \frac{d^3k}{\Omega_{BZ}}|\langle\psi_{i,{\bf k+q}}| {dV_{\rm SCF}\over d {\hat u}_{\nu s} }\cdot
                   \hat \epsilon_{\nu s}|\psi_{j,{\bf k}}\rangle|^2 \times \\
&\times\delta(E_{{\bf k},i} - E_F)  \delta(E_{{\bf k+q},j} - E_F).\\
\end{split}
\end{equation}                 

\noindent
Results are shown in Table \ref{tab:I}.

As $I$, being the electronic part of EPC, measures how the electronic density responds to the atomic vibrations, it is not surprising that the overall $I$ is larger in closed packed metallic Ir than in SrIr$_2$ (4.14 and 2.84 THz$^2$, respectively), where a part of the dense-packed Ir structure is substituted by Sr. 

 \begin{figure}[t]
\includegraphics[width=0.49\textwidth]{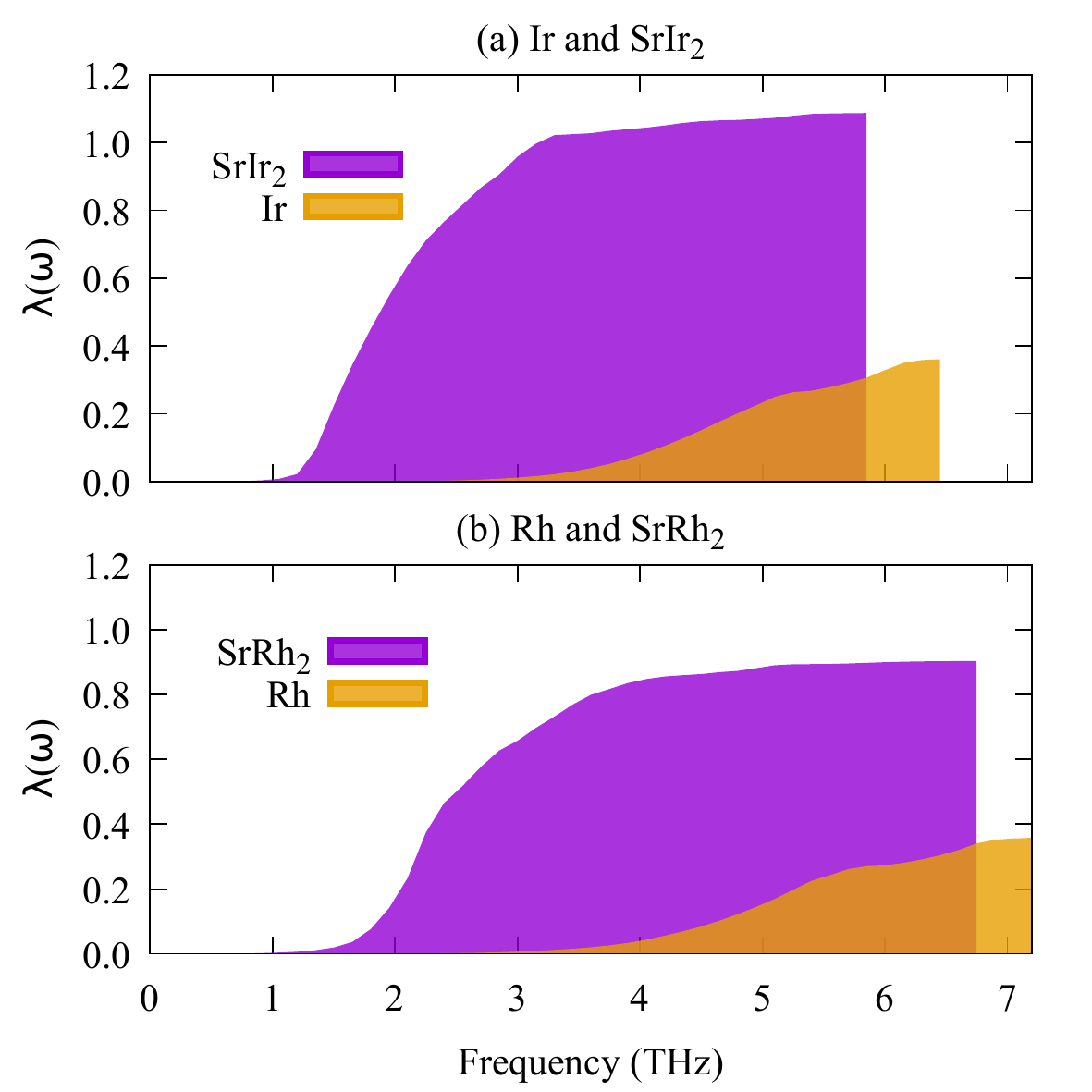}
\caption{$\lambda(\omega) = \int_0^{\omega_{\rm max}} \frac{2}{\omega'}\alpha^2F(\omega'){\rm d}\omega'$: the cumulative frequency distribution of the EPC constant of (a) Ir and SrIr$_2$; and (b) Rh and SrRh$_2$.}\label{fig:lambdy}
\end{figure}

The effect on phonon frequencies, however, outweights  the reduction in the electronic part, 
as the Eliashberg function depends on the ratio $\frac{\gamma_{{\bf q}\nu}}{\omega_{{\bf q}\nu}}$. 
This quantity, as the {\bf q}-dependent contribution to Eliashberg function, is visualized 
in Fig.~\ref{fig:ir-elph}(f,g), and is larger in the case of SrIr$_2$ than Ir. 
In the case of SrIr$_2$ the large contribution comes from ${\bf q} \simeq (0.5, 0, 0)$ (halfway between $\Gamma$ and X). From  Fig.~\ref{fig:ir-elph}(b) we see that it is associated with the softened lowest optical mode, in which Ir atoms move toward the ''empty'' space around Sr, exactly like the previously shown acoustic mode at L-point, which also had a large phonon linewidth. 

 \begin{figure*}[t]
\includegraphics[width=0.80\textwidth]{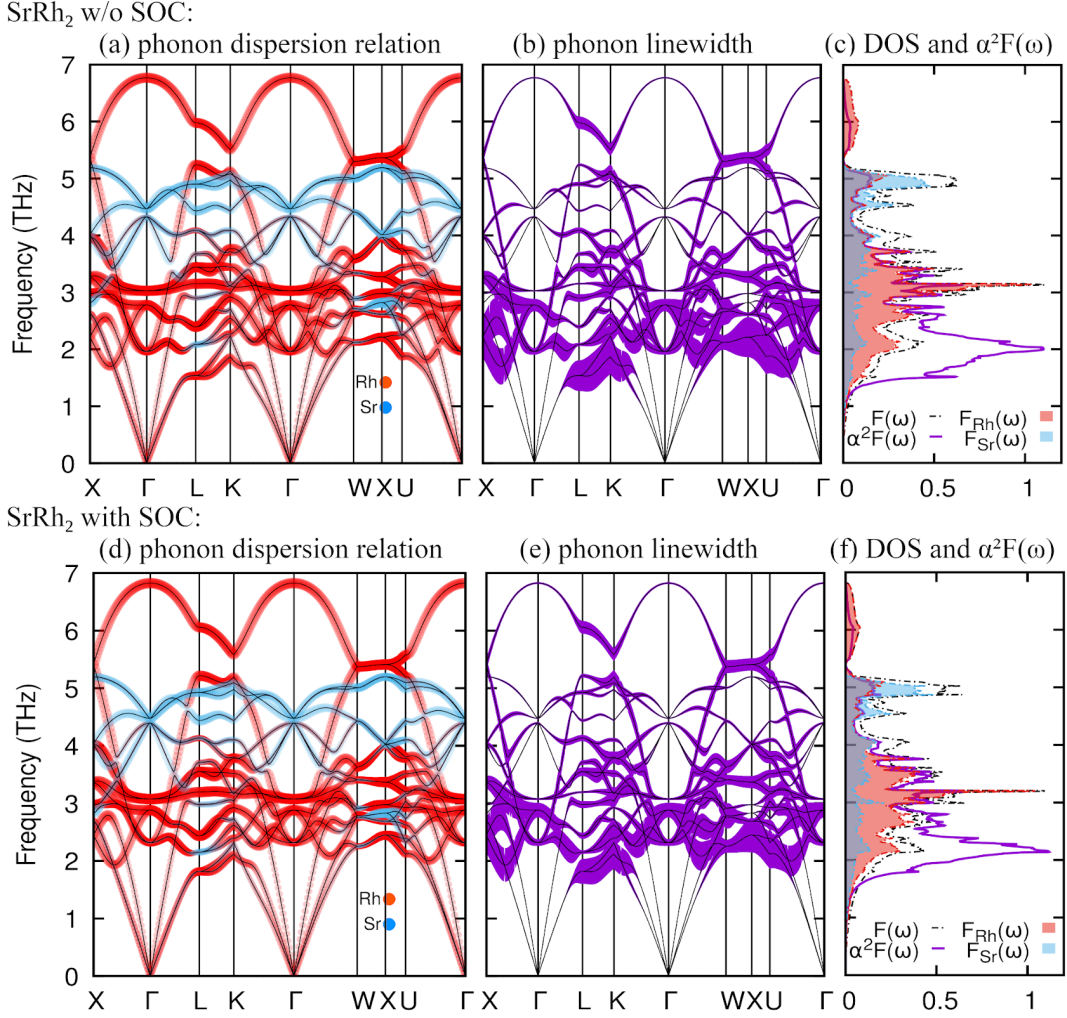}
\caption{The phonon structure and EPC of SrRh$_2$ determined (a-c) without SOC; (d-f)
with SOC; in terms of (a,d) phonon dispersion relation with Sr and Rh contribution marked
with blue and red, respectively; (b,e) with band shading proportional to the phonon linewidths ($\gamma_{{\bf q}\nu}$ in THz multiplied by 4);   (c,f) Eliashberg function and phonon DOS (total and partial); the latter one is normalized as $\alpha^2F(\omega)$.}\label{fig:calc:SrRh2-elph}
\end{figure*}

The lowering of phonon frequencies has an even stronger effect on EPC constant, as $\lambda \propto \frac{\gamma_{{\bf q}\nu}}{\omega^2_{{\bf q}\nu}}$.
In metallic Ir, calculated $\lambda = 0.36$, and is mostly contributed by high-frequency phonons, as shown in Fig.~\ref{fig:lambdy}.
On the other hand, the three-times larger EPC constant of SrIr$_2$, $\lambda =  1.09$, 
is contributed in 94\% by the phonon modes below the pseudogap at 3.40 THz, i.e.
as in metallic Ir is provided by the network of Ir tetrahedrons, but with much reduced phonon frequencies.

In summary, the substitution of Ir tetrahedrons by lighter Sr atoms lowers the phonon frequencies of Ir. That is realized through the less dense atom packing and reduction of the number of nearest neighbors, and turns out to efficiently tune the phonon spectrum to realize the strongly coupled superconducting state.

\subsection{SrRh$_2$}

Similar characteristic of the phonon spectrum and the electron-phonon coupling is found for SrRh$_2$.
In contrast to SrIr$_2$, the phonon spectrum of SrRh$_2$, presented in Fig.~\ref{fig:calc:SrRh2-elph}, is stable in both scalar- and fully relativistic cases, 
showing the lesser importance of SOC for the dynamic properties of this compound, due to the lower atomic number of Rh. 
As Rh is much lighter than Ir ($m_{\text{Rh}} = 102.91$\,u and $m_{\text{Ir}} = 192.22$\,u), the phonon frequencies generally increase: the average $\langle \omega \rangle$, shown in Table \ref{tab:freq}, increases to 3.51 THz. Due to the similar masses of Rh and Sr, we do not observe a pseudo-gap separating the phonon DOS spectrum into two parts. 
However, as far as the atomic character of the phonon modes is concerned, a similar structure as in the previous case is observed. The low-frequency part, up to $\sim 4$ THz, is dominated by Rh with some contribution from Sr, which then dominates between 4 and 5 THz. The characteristic highest bell-shaped optical mode shows up between 5.2 to 6.8 THz. Here it is more separated from the rest of the spectrum and it is also connected to the vibrations of Rh towards the center of the closed-packed Rh tetrahedrons, just like in the analogical mode in SrIr$_2$.

The phonon linewidths have a slightly larger magnitude than in the Ir analog, with the largest associated with the acoustic and the lowest optical modes, as shown in Fig.~\ref{fig:calc:SrRh2-elph}(e). 
This is reflected in the value of the electronic part of EPC, calculated with Eq.(\ref{eq:I}), as $I = 3.46$~THz$^2$ is larger than for SrIr$_2$ (see Table~\ref{tab:I}). 
Modes with large $\gamma_{{\bf q}\nu}$ are concentrated near 2 THz, which produces the peak in the Eliashberg function in Fig.~\ref{fig:calc:SrRh2-elph}(f).
The EPC constant, calculated using Eq.(\ref{eq:lam2}), is equal to $\lambda = 0.90$, lower than for $M$ = Ir in spite of the larger phonon linewidths. 
The reason for this is an increase in the phonon frequencies, and $\lambda(\omega)$ distribution is shifted to higher frequencies, as can be seen in Fig. \ref{fig:lambdy}.
Now, comparing with the values estimated from the experimental measurements, the theoretical result is slightly underestimated with respect to the values of $\lambda = 0.93$, determined from $T_c$, and $\lambda = 1.08$, calculated from the renormalized Sommerfeld parameter. 
All these values indicate slightly weaker electron-phonon coupling than in SrIr$_2$.

\begin{table}[t]
\caption{The electronic part of EPC calculated as an integral $I$ defined in Eq. \ref{eq:I} (expressed in THz$^2$).}
\label{tab:I}
\begin{center}
\begin{ruledtabular}
\begin{tabular}{l c c c c c c}
& 
SrIr$_2$  & 
Ir  &
SrRh$_2$ &
Rh  \\
$I$ w. SOC & 2.84 &
4.14 &
3.46&
4.97\\
\end{tabular}
\end{ruledtabular}
\end{center}
\end{table}

The ''phonon engineering'' effect between Rh and SrRh$_2$ is the same as discussed for SrIr$_2$.
Phonon spectrum of the  metallic Rh is shown in Appendix in Fig.~\ref{fig:ir-rh-elph}, and due to the smaller mass of Rh it is shifted towards higher frequencies when compared to Ir (above 7 THz, in agreement with the experimental data \cite{rh-phonon}), with the calculated EPC constant $\lambda = 0.36$.
The effect of replacement of half of Rh$_4$ tetrahedrons in the Rh 2x2x2 { fcc} supercell with Sr atoms on the phonon dispersion curves is shown in Fig.~\ref{fig:calc:ir-srir2-elph}.
In the same way as in SrIr$_2$, one of the highest Rh phonon modes is recreated in SrRh$_2$ as involving the Rh vibrations towards the center of the remaining Rh tetrahedrons, whereas in the low-frequency part the phonon frequencies are strongly lowered. The average Rh frequency is much reduced, from 5.17 THz in crystalline Rh to 3.40 in SrRh$_2$, see Table~\ref{tab:freq}.
As the crystal locally loses the dense-packed structure, the restoring force acting on a displaced Rh in  SrRh$_2$ is lowered (Fig. \ref{fig:force-constant}) with respect to metallic Rh, while the  bond strength between a pair of Rh atoms remains the same.
Finally, in the context of the electron-phonon coupling, the strong lowering of the phonon frequencies takes over the effect of a moderate reduction in the electronic part of EPC (as shown by the decrease in $I$, see Table~\ref{tab:I}) leading to strong electron-phonon coupling.

\subsection{Effect on the lattice specific heat}

\begin{figure}[b]
\includegraphics[width=0.49\textwidth]{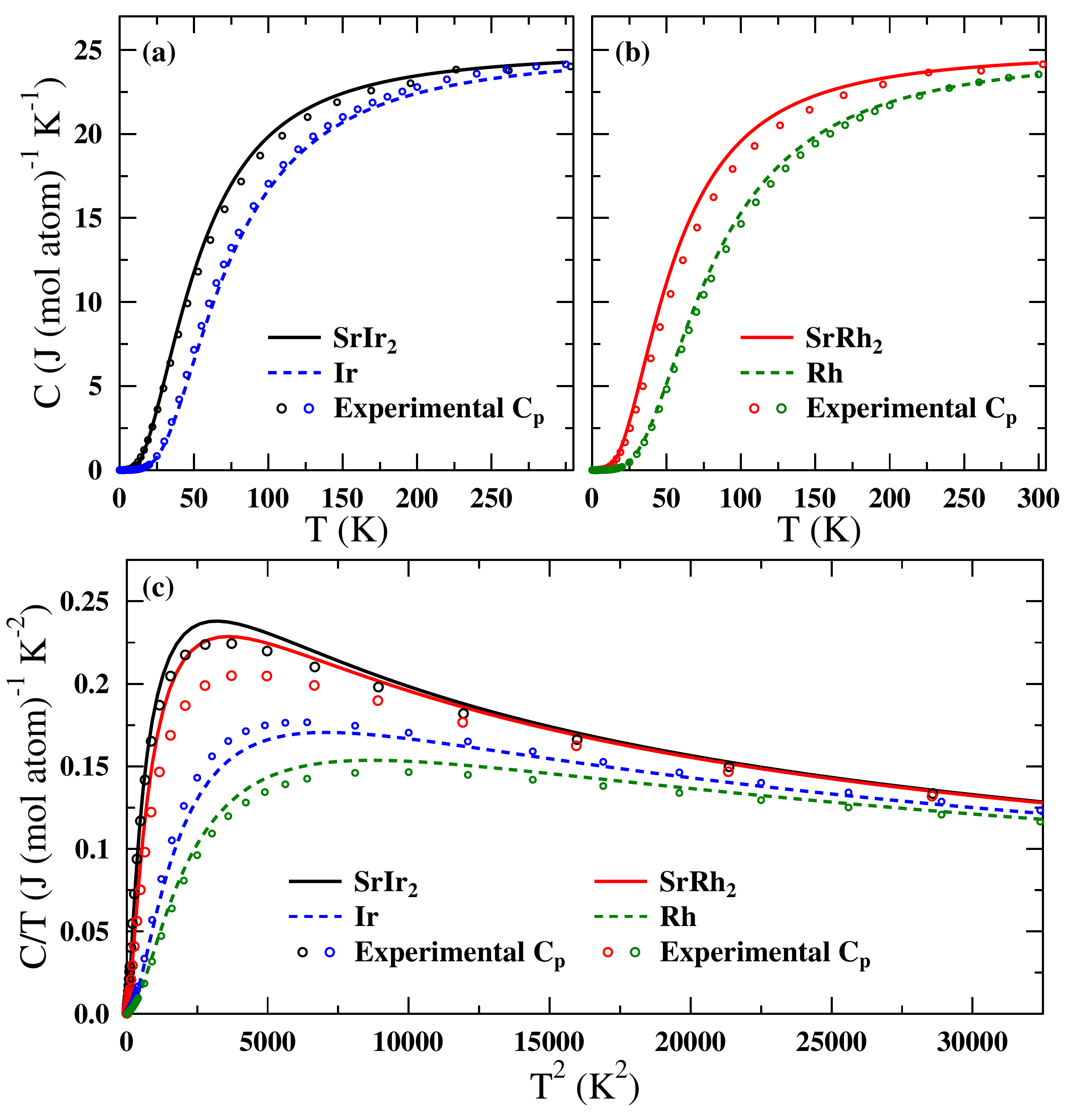}
\caption{Comparison of the lattice specific heat of Sr$M_2$ and $M$, $M$ = Ir and Rh. Lines are the calculated $C_V$, whereas points are $C_p$ taken from our measurements for the Laves phases and Ref.~\cite{heat-metals} for Ir and Rh. The electronic specific heat $\gamma T$ was subtracted from the experimental $C_p$. The transfer of phonons to the lower frequencies is confirmed by a much larger specific heat at lower temperatures. The molar $C_V$ is plotted per atom, to allow for a comparison between the Laves phases and pure metals. }\label{fig:heat-compare}
\end{figure}

Such a strong effect on the phonon spectrum should be visible in the temperature dependence of the heat capacity. 
To confront the results of theoretical phonon calculations with experiment, the constant volume lattice specific heat is calculated as~\cite{grimvall}:
\begin{equation}
C_V = R\int_0^{\infty}F(\omega)\left(\frac{\hbar\omega}{k_BT}\right)^2\frac{\exp(\frac{\hbar\omega}{k_BT})}{(\exp(\frac{\hbar\omega}{k_BT})-1)^2.}
\end{equation}
The phonon density of states $F(\omega)$ functions, shown in Figs.~\ref{fig:calc:SrIr2-elph} and \ref{fig:calc:SrRh2-elph} for Sr$M_2$ and in Fig.~\ref{fig:ir-rh-elph} for $M$, $M$ = Ir and Rh, were used to obtain the theoretical $C_V$ which are plotted in Fig.~\ref{fig:heat-compare}. 
The experimental $C_p$ data (this work for the Laves phases and Ref.~\cite{heat-metals} for Ir and Rh)
are added for comparison.
At first, we see a relatively good agreement between the calculated and experimental results for all presented structures. 
The largest disagreement is seen for SrRh$_2$ on the $C_V/T(T^2)$ plot around the temperature of 60 K, where the calculated $C_V$ is about 8\% larger than the experimental $C_p$. 
This suggests that the calculated phonon spectrum is slightly too soft. 
Nevertheless, both the computed and experimental results confirm our observation of the ''phonon engineering'' effect, which is the transfer of phononic states from higher to lower frequencies, resulting from the structural effect of removing part of the Ir and Rh tetrahedrons while forming the Laves phases. Due to the downward shift of phonon branches, the population of lower-energy phonons, which may be excited at lower temperatures, is increased. 
As a result, the crystal lattice in Sr$M_2$ may absorb larger amounts of thermal energy at lower temperatures, compared to metallic Ir and Rh.
This is confirmed by the increase in $C_V/T$ at lower $T$ in Fig.~\ref{fig:heat-compare}(c) as well as the shift of $C_V(T)$ curve to the left in Figs.~\ref{fig:heat-compare}(a-b).

\section{Superconductivity}

The critical temperatures of the studied compounds are now calculated using the Allen-Dynes formula, Eq.(\ref{eq:tc}),
and presented in Table \ref{tab:tc}.
The Coulomb pseudopotential parameter $\mu^*$ = 0.13 is assumed for all considered compounds, as it is typically used for materials with majority of $d$ states near the Fermi level.

For SrIr$_2$, the logarithmic average frequency is equal to 2.02\,THz (97 K), slightly higher than the  value of 79\,K deduced from the experimental results using the approximate Eq. (\ref{eq:marsiglio}). With the EPC constant 
$\lambda =  1.09$ it gives the critical temperature $T_c = 6.68$\,K, again slightly higher than the experimental value of 6.07\,K, which could be reproduced by taking a larger $\mu^*$ = 0.15.
Nevertheless, the agreement between theory and measurement is very good, and the critical temperature of SrIr$_2$ is dozens of times higher than that of Ir, which is 0.17\,K from our calculations and 0.14\,K from experiment.
As in both cases the superconductivity is driven by the network of Ir tetrahedrons,
this shows the efficiency of the described phonon engineering mechanism to boost the electron-phonon coupling constant in order to achieve superconductivity with a relatively high critical temperature as for conventional systems.
The effect of lowering of the phonon frequencies of Ir reduces the logarithmic average phonon frequency (4.61 THz in metallic Ir versus 2.02 in SrIr$_2$), which is disadvantageous in terms of $T_c$, but the effect of increase in $\lambda$ (from 0.36 to 1.09) compensates for this with an allowance. 

In SrRh$_2$ we see an analogous situation, the logarithmic average frequency is 2.52\,THz (121\,K) and with the EPC constant $\lambda = 0.90$ the calculated critical temperature is $T_c = 5.93$\,K, a bit above the experimental one (5.41\,K).
These values are much enhanced over the calculated in metallic Rh ($\lambda = 0.36$, $T_c = 0.19$~K), which are similar to those obtained in Ir.
However, here the increase in $T_c$ from that measured for metallic Rh is even larger, as 
samples of Rh are almost non-superconducting ($T_c = 0.3$\,mK). 
The reason for the discrepancy of $T_c$ in Rh and such a low experimental critical temperature in rhodium was discussed in Ref. \cite{rhodium-superconductivity} and is due to the presence of spin fluctuations ($\lambda_{\rm sf} \sim 0.1$), revealed also by susceptibility measurements~\cite{rh-susceptibility}. Spin fluctuations compete with superconductivity, effectively lowering the pairing parameter and enhancing the Coulomb repulsion parameter $\mu^*$. This  significantly lowers $T_c$ of Rh below the value in Ir, in spite of the similar EPC constants $\lambda$.
Such effect is not seen in the SrRh$_2$ case. 

It is worth to comment here on the spin-orbit coupling effect on the superconductivity in the studied Laves phases. 
As for the SrIr$_2$ the phonon structure was unstable in the scalar-relativistic case, we can only comment it for SrRh$_2$. 
There, the scalar-relativistic values of the logarithmic average frequency and EPC constant are $\langle\omega_{\rm log}^{\alpha^2F}\rangle=2.24$\,THz and $\lambda=1.12$, respectively. The resulting  critical temperature is 8.00\,K, larger than in the fully-relativistic case due to the larger $\lambda$. 
Thus, the spin-orbit coupling is not beneficial for superconductivity in this case. Similar situation was previously found in CaIr$_2$ and CaRh$_2$ Laves phases~\cite{cair2-carh2-tutuncu} as well as in CaBi$_2$~\cite{cabi2}, which is a distorted Laves phase.

\section{Validation of the computed Eliashberg functions}

The measured temperature dependence of the specific heat, discussed at the beginning of this work in Fig.~\ref{fig:exp:heat-capacity}, allows us to more precisely validate the accuracy of the computed Eliashberg functions than by comparing single parameters, like $\lambda$ and $T_c$. 
By using the Eliashberg gap equations (for more details see~\cite{thcoc2,Eliashberg1960,Broyden1984,Carbotte1990} and the Supplemental Material \cite{supplemental}) the temperature dependence of the electronic specific heat in the superconducting state, as well as the specific heat jump at the transition temperature may be calculated from the $\alpha^2F(\omega)$ functions with just one external parameter, the Coulomb pseudopotential. 
For such calculations, the ''computational'' value of $\mu^*_c$ is determined in a first step to obtain the superconducting critical temperature equal to the experimental one and then thermodynamic quantities are computed as a function of $T/T_c$. The $\mu^*_c$ obtained from Eliashberg equations is usually different (larger) than the value of $\mu^*$ which reproduces experimental $T_c$ in the Allen-Dynes formula~\cite{thcoc2,szczesniak_2015,Morel1962,mu_star},  that's why we have used a different symbol. The reason behind this difference is discussed in Ref.~\cite{allen-dynes} and is related to the dependence of $\mu^*_c$ on the cutoff frequency, used while solving the Eliashberg equations. 
In our case $\mu^*_c$ is $0.261$ (SrIr$_2$) and $0.243$ (SrRh$_2$), and to be compared with previously used $\mu^*$ has to be re-scaled (more details in Supplemental Material \cite{supplemental}) resulting in  $\mu^* = 0.169$ and 0.161, respectively. These re-scaled values are now close to $\mu^* \simeq 0.15$, required to reproduce the experimental $T_c$ using Allen-Dynes formula. This small enhancement of  $\mu^*$ over the conventional $(0.10 - 0.13)$ range of values may be related to spin-orbit coupling effects, Fermi surface complexity and slight enhancement of electronic interactions, as can be expected for our $d$-band metals.

\begin{figure}[t]
\includegraphics[width=0.49\textwidth]{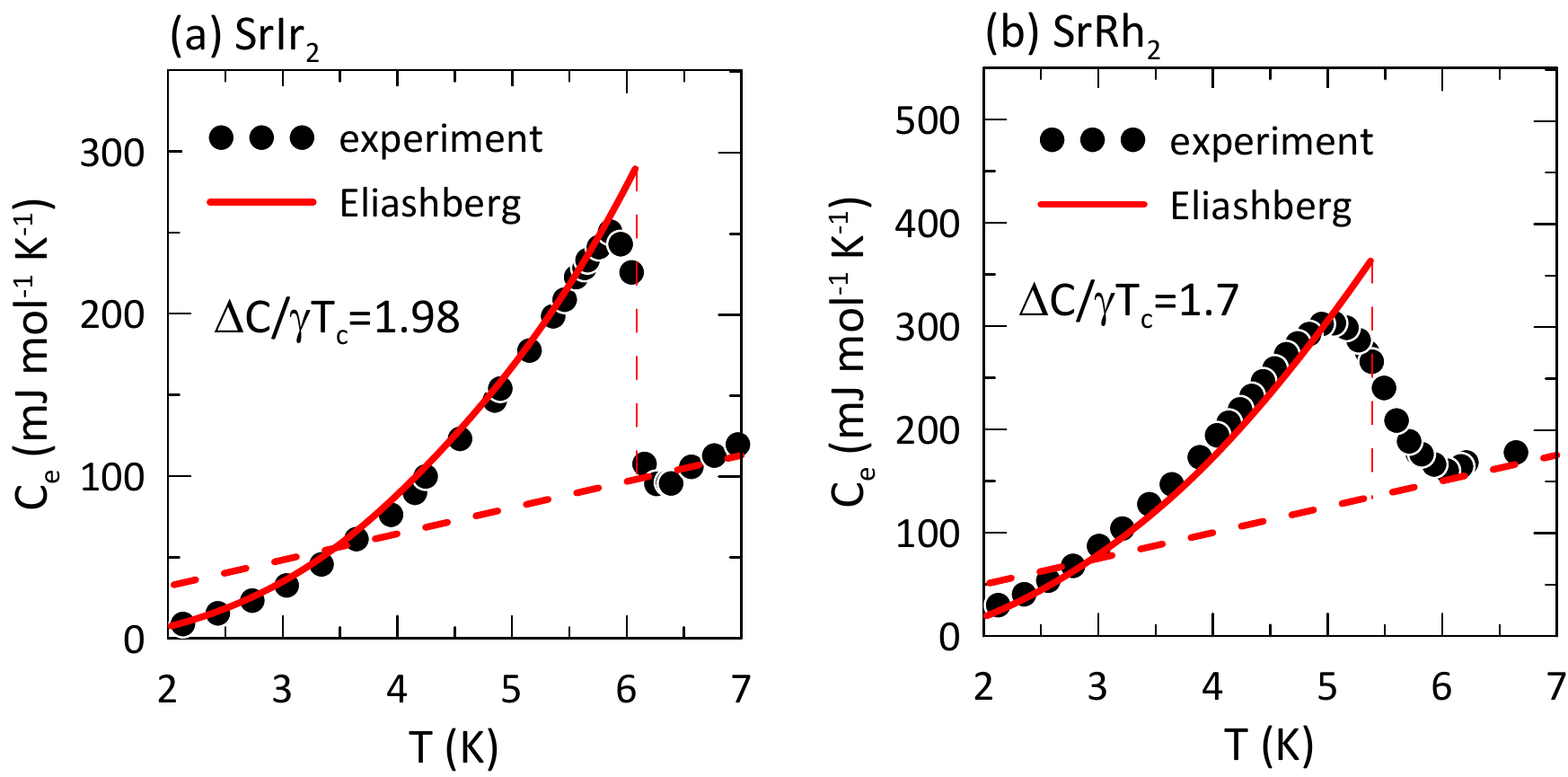}
\caption{Calculated and experimental electronic specific heat in (a) SrIr$_2$; and  (b) SrRh$_2$. The computed reduced specific heat jump at the superconducting critical temperature is shown in the figures.}\label{fig:heat}
\end{figure}

The calculated temperature dependencies of electronic heat capacity $C_e$ are shown in Fig.~\ref{fig:heat}.
The agreement between the computed and experimental results is worth emphasizing. The temperature profile of $C_e(T)$ and the value of the measured specific heat jump at the superconducting transition temperature remain in a very good agreement, the computed versus measured $\Delta C/\gamma T_c$ are 1.98 versus 2.08 for SrIr$_2$ and 1.70 versus 1.80 for SrRh$_2$. 
This finally confirms the strong electron-phonon coupling character of Sr$M_2$ superconductors which are very well described by the isotropic $s-$wave Eliashberg theory. 

\section{Summary and conclusions}
In summary, the electronic structure, lattice dynamics, and superconductivity of two cubic Laves phase superconductors, Sr$M_2$, $M =$ Ir and Rh, $T_c = 6.07$\,K and 5.41\,K, were studied theoretically and experimentally.  
The crystal structure of both compounds is based on the underlying { fcc} structures of metallic Ir and Rh, where half of the atomic $M_4$ tetrahedrons are substituted by Sr atoms.
From the electronic structure point of view, near the Fermi level, Sr  effectively acts as an electron donor, shifting $E_F$ above the characteristic peak in the density of states seen also in the case of elemental $M$ metallic structures. 
More significant changes are made to the phonon structure, where 
locally, after $M_4 \rightarrow$ Sr substitution, the structure looses the dense-packed character.
As a consequence, the frequencies of some of the Ir/Rh phonon modes, propagating in the network of tetrahedrons now spaced by Sr atoms, are reduced twice, compared to the elemental metal cases. 
This transfer of phononic states to the low frequency range is confirmed by the enhanced low-temperature specific heat and results in the strongly enhanced electron-phonon coupling.
The enhancement in $\lambda$ is possible  
because the electronic contribution to the EPC is less reduced in comparison to the average square Ir phonon frequency. 
Finally, the very-low-$T_c$, weakly-coupled superconductors Ir and Rh ($\lambda \sim 0.35$, $T_c = 0.14$ and $0.3\times 10^{-3}$ K) become strongly coupled Laves phases, with $\lambda \sim  1$, the higher as the larger is the mass of $M$ and the lower their phonon frequencies.
This results in much higher critical temperature values of $T_c = 6.07$ and $5.41$ K in SrIr$_2$ and SrRh$_2$, respectively.

The superconducting properties of Sr$M_2$ may be compared to a related isoelectronic family of Laves phases Ca$M_2$ ($M =$ Ir, Rh) \cite{haldo-cair2,carh2-gornicka,cair2-carh2-tutuncu}, which superconduct with slightly lower critical temperatures of 5.8\,K and 5.1\,K.
As they have the same crystal structure, the ''phonon engineering'' mechanism should also be responsible for the strongly-coupled superconductivity in Ca$M_2$, where $\lambda \simeq 1$~\cite{cair2-carh2-tutuncu}. Analyzing the theoretical phonon and electron-phonon calculations reported in Ref.~\cite{cair2-carh2-tutuncu} indeed we see that the same mechanism may be identified. 
The electron-phonon coupling is dominated by Ir/Rh vibrations, whose frequencies are much reduced from their elemental metallic values after $M_4 \rightarrow$ Ca substitution. 
Moreover, the smaller mass of Ca compared to Sr is correlated with the lower $T_c$.
This trend is further confirmed by the data obtained for BaRh$_2$ \cite{gong-srrh2-barh2}, which has the highest critical temperature (5.6\,K) among all three $X$Rh$_2$ compounds ($X =$ Ca, Sr, Ba). 

Summarizing this comparison, two ways of increasing the critical temperature among $XM_2$ Laves phases family of compounds are observed: when we substitute $M$ with the isoelectronic heavier element from the next group of periodic table (Rh $\rightarrow$ Ir), then $T_c$ is increased by $\sim 10$\%. Analogously, when $X$ atom is substituted with a heavier element (Ca $\rightarrow$ Sr $\rightarrow$ Ba), then $T_c$ is also increased, but the effect is weaker ($T_c$ is changed of about 5\%)~\footnote{Interestingly, all these trends are opposite to the one arising from the very first works of Matthias \cite{matthias} ($T_c$=5.7\,K for SrIr$_2$, 6.2\,K for SrRh$_2$, 4-6.15\,K for CaIr$_2$, 6.4\,K for CaRh$_2$ and 6.0\,K for BaRh$_2$).}. 
The basis for superconductivity in all mentioned $XM_2$ is the significant reduction of phonon frequencies of $M$ transition metals due to the local dilution of their closed-packed crystal structure.
The understanding of this phonon engineering mechanism may potentially help to design new strongly coupled electron-phonon superconductors.

\section{Acknowledgments}
This work was supported by the National Science Centre (Poland), grant numbers 2017/26/E/ST3/00119 (BW and SG), 
2017/26/D/ST3/00109 (PW) and UMO-2019/33/N/ST5/01496 (KG and TK).
SG was also partly supported by the EU Project POWR.03.02.00-00-I004/16.

\section*{Appendix}
\begin{figure*}[htb!]
\includegraphics[width=1.00\textwidth]{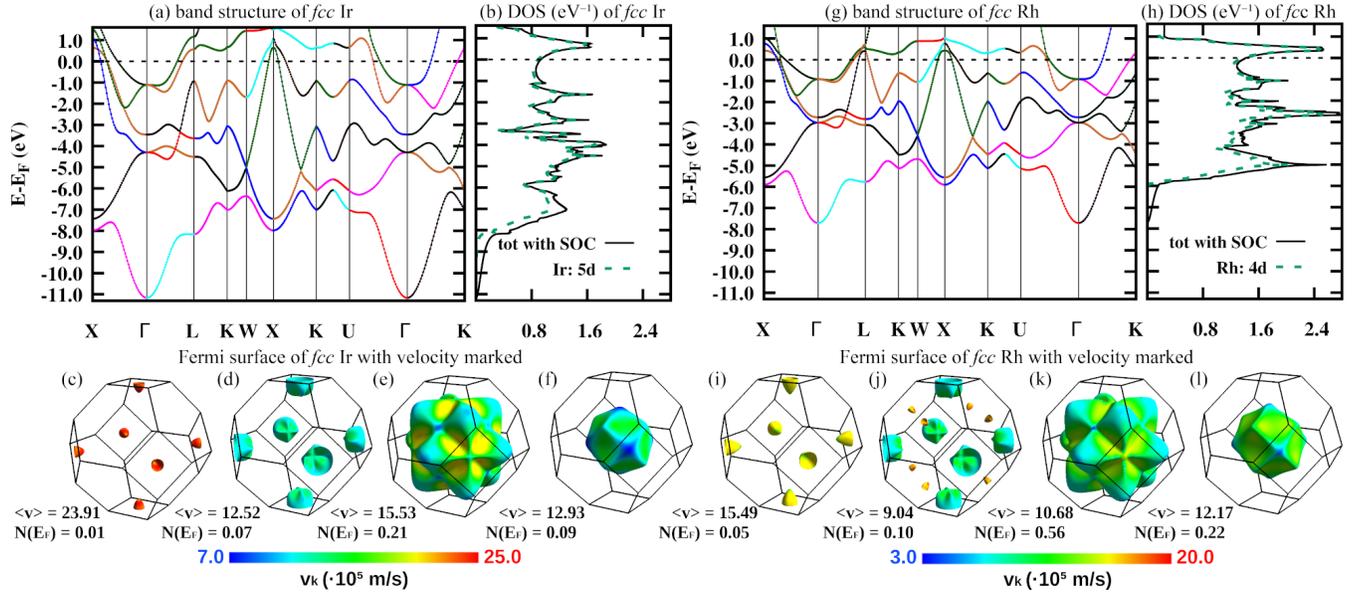}
\caption{Electronic band structure of (a-f) Ir;  and (g-l) Rh.
Band structure (the points matched to each other with help of symmetry analysis are connected with
colored lines), DOS with atomic contributions and Fermi surface colored with respect to Fermi velocity. Average Fermi velocity $\langle v \rangle$ ($10^5$~m/s) and density of states $N(E_F)$ (eV$^{-1}$) for each of the FS sheets are shown.
}\label{fig:ir-rh-el}
\end{figure*}

\begin{figure*}[htb!]
\includegraphics[width=1.00\textwidth]{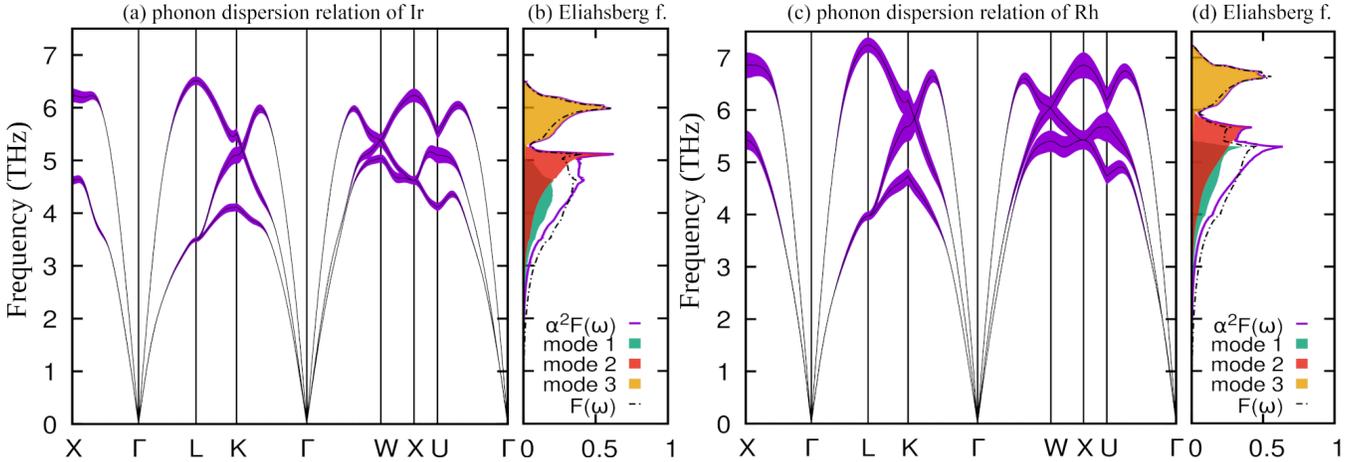}
\caption{Phonon dispersion relations with band shading 
proportional to the phonon linewidths ($\gamma_{{\bf q}\nu}$ in THz multiplied by 4) of (a) Ir; and (c) Rh. Electron-phonon interaction function $\alpha^2F(\omega)$ and phonon DOS $F(\omega)$ normalized to $\alpha^2F(\omega)$ for (b) Ir; and (d) Rh.}\label{fig:ir-rh-elph}
\end{figure*}

\appendix
Electronic structures of crystalline metallic { fcc} Ir and Rh are shown in Fig.~\ref{fig:ir-rh-el}. Phonon dispersion relations, phonon densities of states $F(\omega)$ and the electron-phonon interaction functions $\alpha^2F(\omega)$ are shown in Fig.~\ref{fig:ir-rh-elph}. 

\newpage
\bibliography{references}

\newpage

\onecolumngrid

\section*{Supplemental Material}

\renewcommand{\thefigure}{{S\arabic{figure}}}
\setcounter{figure} 0

\section{X-ray powder diffraction}
\begin{figure}[htb!]
\includegraphics[width=0.70\textwidth]{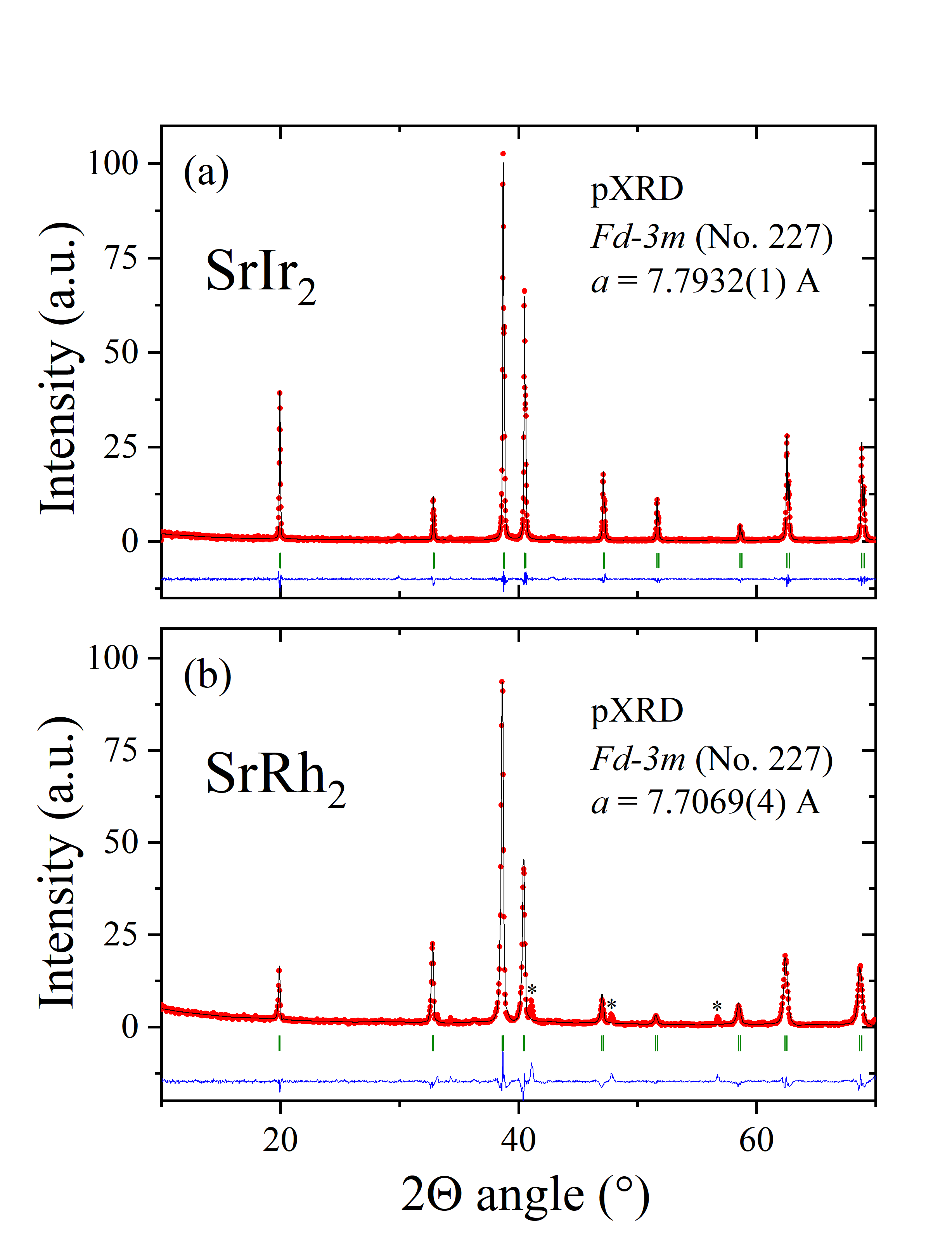}
\caption{The room-temperature X-ray powder diffraction (pXRD) patterns together with the LeBail refinements of SrIr$_2$ and SrRh$_2$. Our analysis confirms that both compounds crystallize in a cubic crystal structure (space group Fd-3m, No. 227). The pXRD pattern for SrRh$_2$ indicates that a sample is nearly single phase with small amount of impurity (denoted by asterisks). No impurities  were detected in a sample containing Ir. The lattice parameters obtained from the LeBail refinement are a = 7.7932(1) Å and a = 7.7069(4) Å for SrIr$_2$ and SrRh$_2$, respectively. These values are in a good agreement with the data reported previously in Ref. Wood et al. [25], Horie 
et al. [12], Gong et al. [13]. The difference plot (between experimental and fitted data) and the expected Bragg peak positions are also presented.}
\label{fig:xrd}
\end{figure}

\section{Phonon modes of Ir supercell}
\begin{figure}[h]
\includegraphics[width=0.75\textwidth]{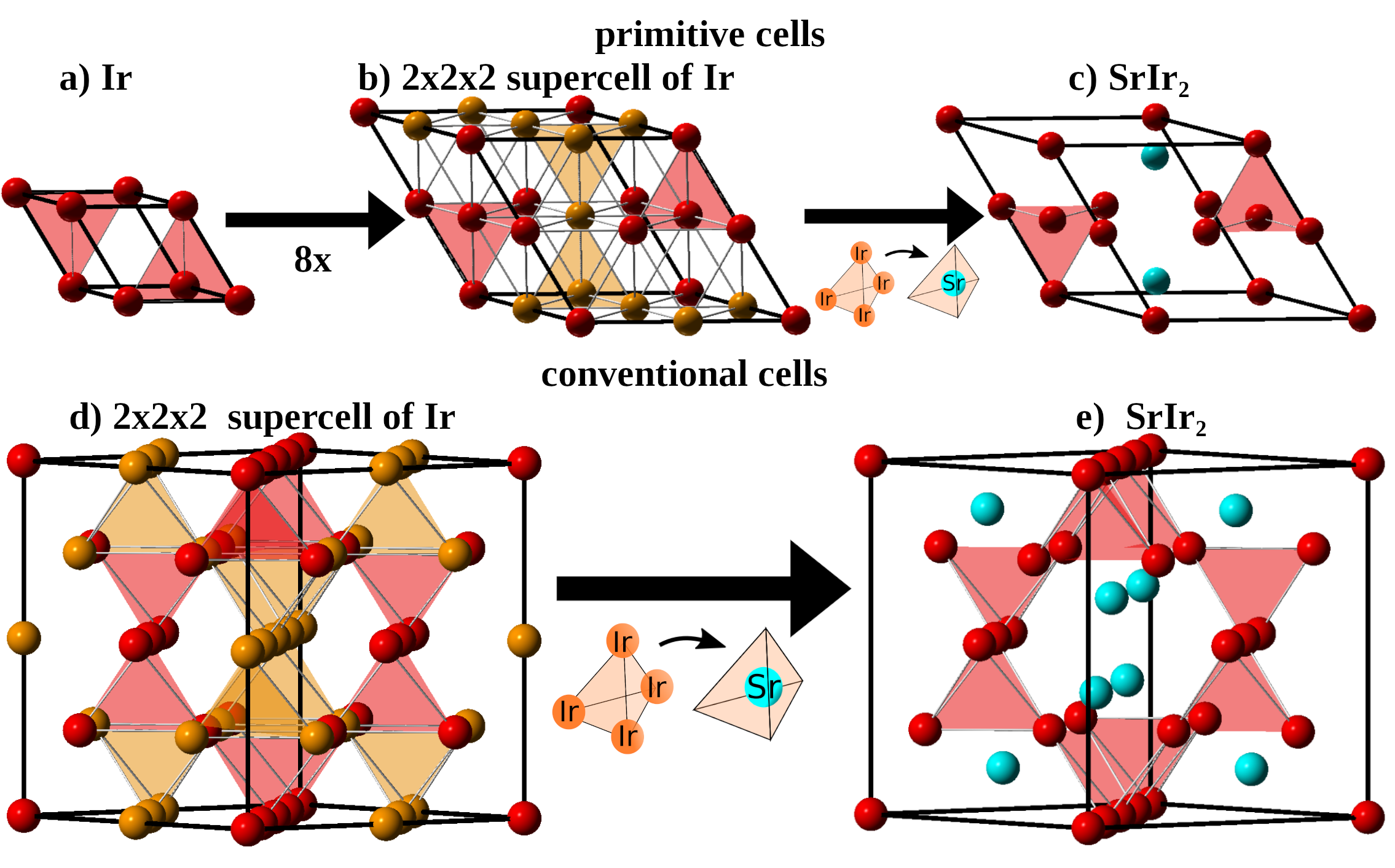}
\caption{ The structure of $fcc$ Ir (a), 2x2x2 supercell of Ir (b) and of SrIr$_2$ (c), which can be obtained by substitution of part of tetrahedrons made of Ir atoms by Sr atoms. The tetrahedrons (and Ir atoms) marked with orange color in supercell of Ir are replaced by Sr in SrIr$_2$ structure, while red tetrahedrons (and Ir atoms) remain the same. The panels (a-c) show primitive cell of these structures, while in panels (d-e) the conventional supercell of Ir and unit cell of SrIr$_2$ are shown. It shall be stressed, that Ir structure consists of closed packed tetrahedrons (there is no empty space between tetrahedrons), however the ones which consist of both red and orange Ir atoms are omitted on panels (a, d) for clarity of the figure. }
\label{fig:structure2}
\end{figure}

The crystal structure of Ir is made of Ir tetrahedrons, while the crystal structure of SrIr$_2$ can be seen as a supercell of Ir with half of tetrahedrons replaced by Sr (see Fig. \ref{fig:structure2}). In the main text we explain how it is connected to a phonon structure. Particularly, we have shown that the highest mode of SrIr$_2$  is associated with a movement of Ir toward the center of tetrahedron. In the case of Ir supercell, similarly shaped, high frequency modes are present and here we would like to describe how they are different from the corresponding SrIr$_2$ mode.\\

\begin{figure}[h]
\includegraphics[width=1.00\textwidth]{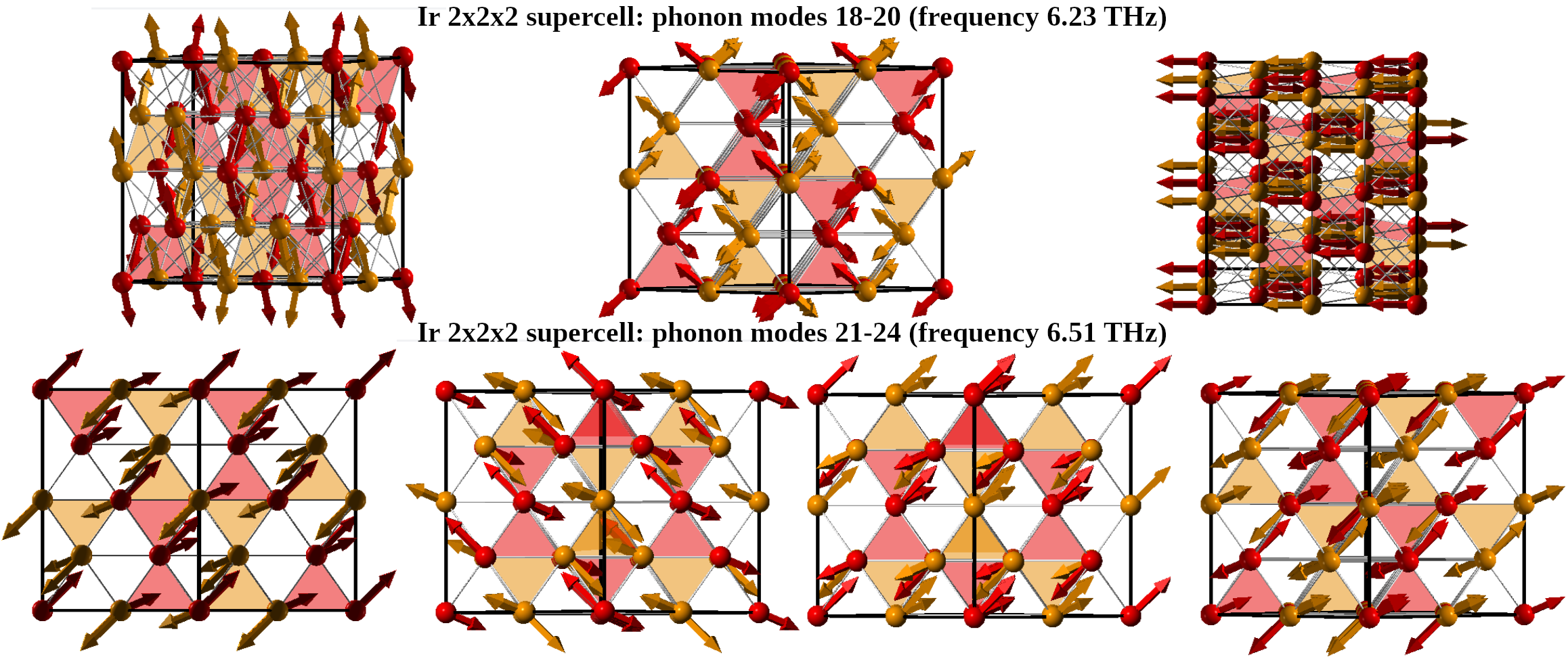}
\caption{ The seven highest phonon modes of 2x2x2 supercell of Ir shown in real space with displacement vectors marked with arrows. }
\label{fig:phonons}
\end{figure}

In Fig. \ref{fig:phonons} the seven highest phonon modes of Ir 2x2x2 supercell at $\bm{q}=(0, 0, 0)$ are shown in real space as arrows attached to the atoms. Three of them are degenerated at a frequency equal to $6.23$\,THz, while other four are degenerated at $\omega=6.51$\,THz. These Ir atoms and tetrahedrons which are replaced by Sr in SrIr$_2$ structure are colored with orange, while the those which survive in SrIr$_2$ are marked with with red color.\\
The modes at $6.23$\,THz are associated with a movement of Ir atoms in $XZ$ (along Ir-Ir bondings) or $XY$ plane (along $x$ axis). The modes at $6.51$\, THz are of more complicated nature. However, all of them lead to a deformation of tetrahedrons in a way that one of the four atoms, which form tetrahedron, is moving to the center of a face of tetrahedron.\\
In all the cases, among the four atoms, which form the tetrahedron, two pairs can be distinguished such that within one pair atoms are moving approximately along one axis (in opposite directions), so the sum of polarization vectors of atoms belonging to one tetrahedron is equal to zero, as expected, since whole structure is built of tetrahedrons and a sum over unit cell of all polarization vectors of optical mode has to be zero. In the case of SrIr$_2$ it is not the case, since half of tetrahedrons are replaced by Sr, thus all of atoms within one tetrahedron can move toward its center.

\section{Eliashberg formalism and specific heat calculations}
The specific heat calculations are based on the isotropic Eliashberg equations~(Eliashberg [47], Grimvall [34]) which, on the imaginary axis, can be written as
\begin{eqnarray}
Z(i\omega _n)&=&1+\frac{\pi k_B T}{\omega _n} \sum _{n'} \frac{\omega _{n'}}{R(i\omega _{n'})} \lambda(n-n'),
 \label{ee1} \\
 Z(i\omega _n) \Delta(i\omega _n) &=& \pi k_B T \sum _{n'} \frac{\Delta(i\omega _{n'})}{R(i\omega _{n'})} 
 [ \lambda(n-n')-\mu^*_c \theta ( \omega _c - \omega _{n'} )],
 \label{ee2}
\end{eqnarray}
where $Z(i\omega_n)$ is the mass renormalization function, $\Delta(i\omega _n)$ is the superconducting order parameter, $i\omega_n=i(2n+1)\pi k_B T$ 
are fermionic Matsubara frequencies where $n\in \mathbb{Z}$, $\theta(\omega)$ is the Heviside function, $k_B$ is the Boltzmann constant, $T$ is the
temperature and $R(i\omega _n) = \sqrt{\omega _n ^2 + \Delta^2(i\omega _n)}$. 
The Coulomb pseudopotential $\mu^*_c$ is determined by the electron-electron interactions and usually takes the value in the range $[0.1,0.2]$~(Carbotte [49]). Note, however, that when the Eliashberg equations are solved, $\mu^*_c$ depends on the cutoff frequency and usually larger value of  $\mu^*$ is required to obtain the experimental value of $T_c$, than when Allen-Dynes or McMillan formulas are used. 
To properly determine the thermodynamic properties of the studied superconductor, the value of $\mu^*_c$ has to be adjusted to match the calculated and experimental $T_c$, {and should be rescaled to be compared to the usual $\mu^*$ (see below).}

\begin{figure}[b]
\includegraphics[scale=.5]{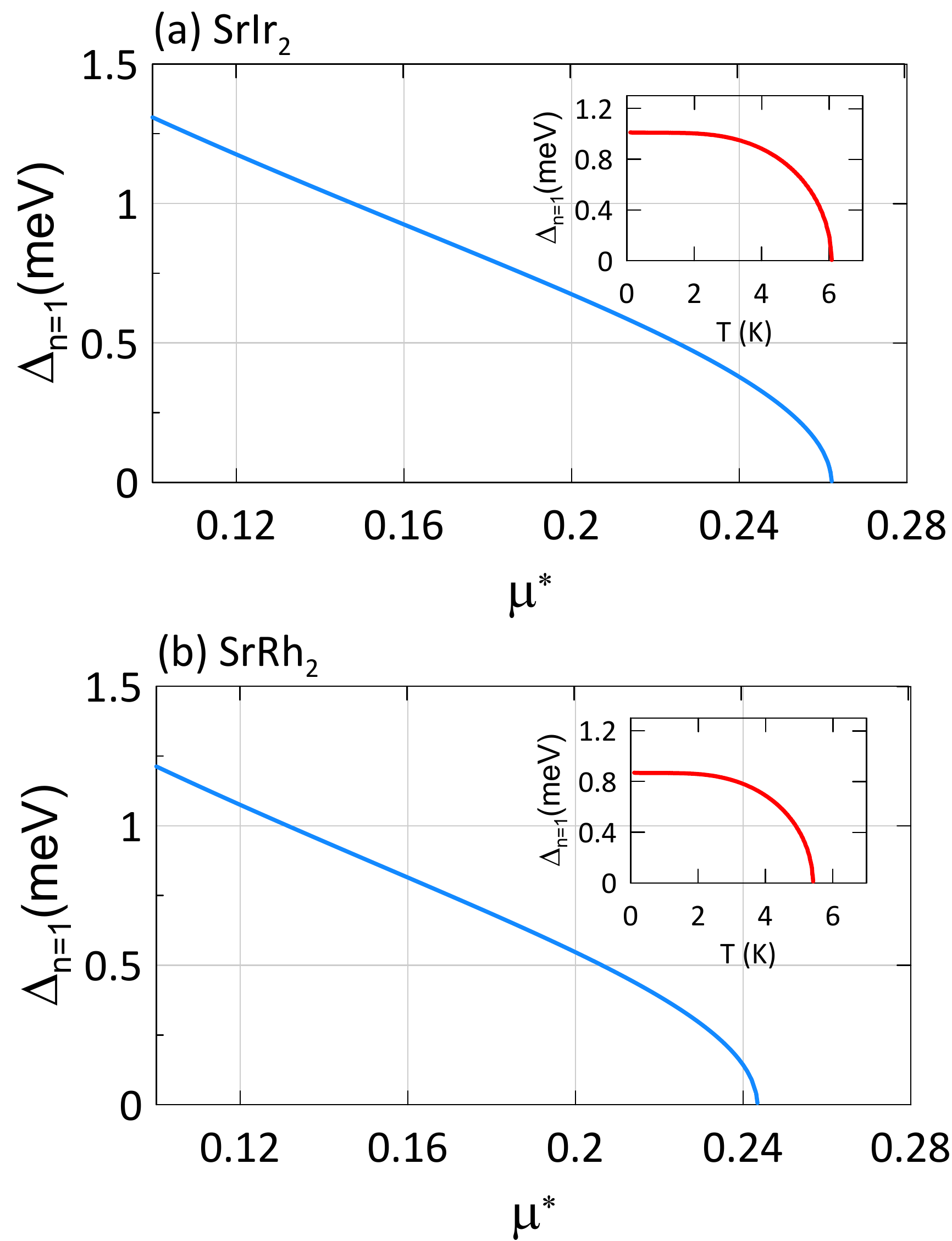}
\caption{ The superconducting energy gap $\Delta_{n=1}$ evaluated at $T=6.07$~K (SrIr$_2$) and  $T=5.4$~K (SrRh$_2$), as a function of $\mu^*_c$. Insets 
presents $\Delta_{n=1}(T)$ for $\mu^*_c$ for which the calculated $T_c$ correspond to experimental ones. }
\label{fig:muTc}
\end{figure}

The kernel of the electron-phonon interaction 
\begin{equation}
 \lambda(n-n')=\int _0^{\infty} d \omega \frac{2 \omega \alpha ^2 F(\omega)}{(\omega _n - \omega _{n'})^2+\omega ^2},
\end{equation}
where $\alpha ^2 F(\omega)$ is the isotropic Eliashberg spectral function.\\
The isotropic Eliashberg equations (\ref{ee1})-(\ref{ee2}) are solved iteratively until the convergence is reached, which we consider to occur when 
the relative variation of $\Delta(i\omega_n)$ between two consecutive iterations is lower than $10^{-15}$. The number of iterations is reduced by  the use of the Broyden method to predict subsequent solutions~(Srivastava [48]). The calculations are performed for the cut-off frequency 
$\omega _c = 8 \omega _{\rm max}$~(Grimvall [34]) and the number of Matsubara frequencies $M=6000$. The self-consistent solution of 
Eqs.~(\ref{ee1})-(\ref{ee2}) for a given Eliashberg spectral function $\alpha^2F(\omega)$ is then used to calculate  the energy difference between the superconducting and normal state $\Delta F$, which is given by
\begin{eqnarray}
 \frac{\Delta F}{N(E_F)}&=&-\pi k_BT \sum _{n} \left ( \sqrt{\omega_n^2+\Delta _n^2} - |\omega _n|\right ) 
 \left ( Z^S(i\omega _n) - Z^N(i\omega _n) \frac{|\omega _n|}{\sqrt{\omega _n^2+\Delta_n^2}} \right ),
\end{eqnarray}
where $N(E_F)$ corresponds to the density of states at the Fermi level while $Z^S$ and $Z^N$ denote the mass renormalization factors 
for the superconducting (S) and normal (N) states, respectively. \\
Finally, the difference in the electronic specific heat $\Delta C_e=C_e^S-C_e^N$ can be expressed as
\begin{equation}
 \frac{\Delta C_e(T)}{k_B N(E_F)} = -\frac{1}{\beta}\frac{d^2 \Delta F /N(E_F)}{d(k_BT)^2},
\end{equation}
with the specific heat in the normal state given by
\begin{equation}
\frac{C_e^N(T)}{k_B N(E_F)}=\frac{\pi^2}{3} k_B T (1+\lambda),
\end{equation}
where $\lambda$ is the electron-phonon coupling constant.\\

As mentioned above, we use the common practice in which $\mu^*_c$ is determined based on the experimental value of $T_c$.
For this purpose, we calculated $\Delta_{n=1}(T=T_c)$ for different $\mu^*_c$, see  Fig.~\ref{fig:muTc}. The value of $\mu^*_c$ for which $\Delta_{n=1}(T=T_c)=0$ is then used to  evaluate the specific heat of the considered compounds. Based on Fig.~\ref{fig:muTc} we can determine $\mu^*_c=0.261$ for SrIr$_2$ and $\mu^*_c=0.243$ for SrRh$_2$.

The temperature dependencies of $\Delta_{n=1}$ for determined $\mu^{*}_c$ are presented in 
insets of Fig.~\ref{fig:muTc} and undergoes the following formula
\begin{equation}
 \Delta(T)=\Delta(0)\sqrt{1-\left ( \frac{T}{T_c} \right )^\Gamma},
 \label{eq:delT}
\end{equation}
with $\Gamma=3.38$ $(3.33)$ for SrIr$_2$ (SrRh$_2$) slightly larger than predicted from the BCS theory, $\Gamma_{BCS} \approx 3.0$. The extrapolated 
$\Delta(0)=1.01$ $(0.87)$~meV gives the dimensionless ratio $R_{\Delta} = 2 \Delta (0) / k_B T_c=3.86$ $(3.73)$ close to the BCS value $3.53$. 
{Following Allen and Dynes [31], due to the cut-off frequency dependence, the ''numerical'' $\mu^*_c$ has to be scaled to be compared to the conventional $\mu^*$ according to the formula: 
\begin{equation}
\frac{1}{\mu^*} = \frac{1}{\mu^*_c} + \ln\left(\frac{\omega_c}{\omega_{\rm max}}\right).
\end{equation}
In our case ${\omega_c}/{\omega_{\rm max}} = 8$, and re-scaled $\mu^* = 0.169$ (SrIr$_2$) and $\mu^* = 0.161$ (SrRh$_2$). These re-scaled values are now close to $\mu^* \simeq 0.15$, required to reproduce the experimental $T_c$ in SrIr$_2$ and SrRh$_2$ using the Allen-Dynes formula. This small enhancement of  $\mu^*$ over the conventional (0.10 - 0.13) range of values may be related to spin-orbit coupling effects, Fermi surface complexity and slight enhancement of electronic interactions, as can be expected for the $d$-band metals. 
}

\end{document}